\documentclass[a4paper,reprint,twocolumn,notitlepage,aip,cha,nofootinbib]{revtex4-2}
\usepackage[left=1.5cm,right=1.5cm,top=1cm,bottom=1.5cm,includeheadfoot]{geometry}
\usepackage[english]{babel}
\usepackage[T1]{fontenc}
\usepackage{tgtermes} 
\usepackage{tgheros} 

\usepackage{amsmath}
\usepackage{amssymb}
\usepackage{amsthm}
\usepackage{mathtools}
\usepackage{bm}
\usepackage{accents}
\usepackage{stmaryrd}

\usepackage{graphicx}
\usepackage[dvipsnames]{xcolor}
\usepackage{tikz}
\usepackage{pgfplots} 
\usetikzlibrary{shapes,shadings,positioning}

\usepackage{soul}
\usepackage{cancel}
\usepackage{framed}

\PassOptionsToPackage{hyphens}{url} 
\usepackage[breaklinks=true]{hyperref}
\bibpunct{[}{]}{;}{n}{}{} 

\usepackage[nohyperlinks, printonlyused, withpage, smaller]{acronym} 

\usepackage{enumerate}
\usepackage[shortlabels]{enumitem}

\theoremstyle{remark}

\theoremstyle{definition}

\newcommand{\rmi}{\mathrm{i}}
\newcommand{\rme}{\mathrm{e}}

\newcommand{\rmd}{\mathrm{d}}
\newcommand{\rmm}{\mathrm{m}}
\newcommand{\R}{\mathbb{R}}

\newcommand{\eps}{\varepsilon}
\newcommand{\subdiff}{\underline\partial}
\newcommand{\supdiff}{\overline\partial}
\DeclareMathOperator{\trace}{Tr}

\newcommand{\weakto}{\rightharpoonup}
\newcommand{\Lloc}{L^1_\mathrm{loc}}

\DeclareRobustCommand\bar[1]{\accentset{\rule{.4em}{.3pt}}{#1}}

\newcommand{\rr}{\bm{r}}
\newcommand{\kk}{\bm{k}}
\renewcommand{\aa}{\bm{a}}
\renewcommand{\AA}{\bm{A}}
\newcommand{\jj}{\bm{j}}
\newcommand{\hAA}{\skew{5}\hat{\AA}} 
\newcommand{\bAA}{\skew{5}\bar{\AA}} 
\newcommand{\bjj}{\skew{5}\bar{\jj}} 
\newcommand{\EE}{\bm{E}}
\newcommand{\BB}{\bm{B}}
\newcommand{\DD}{\bm{D}}
\newcommand{\HH}{\bm{H}}
\newcommand{\PP}{\bm{P}}
\newcommand{\MM}{\bm{M}}

\newcommand{\jjb}{\bm{j}_\mathrm{b}}
\newcommand{\jjf}{\bm{j}_\mathrm{f}}
\newcommand{\JJ}{\bm{J}}
\newcommand{\hJJ}{\skew{5}\hat{\JJ}} 
\newcommand{\mm}{\bm{m}}
\newcommand{\be}{\bm{\epsilon}}
\newcommand{\wC}{w_\mathrm{C}}
\newcommand{\Fren}{\bar F_\mathrm{ren}}
\newcommand{\EHxc}{E_\mathrm{Hxc}}
\newcommand{\EMregHxc}{\bar E_{\mathrm{Hxc}}}
\newcommand{\phiconf}{\phi_\mathrm{conf}} 

\newcommand{\phis}{\phi_\mathrm{s}}

\newcommand{\phiHxc}{\phi_\mathrm{Hxc}}
\newcommand{\phiMregHxc}{\bar\phi_{\mathrm{Hxc}}}

\newcommand{\rhos}{\rho_\mathrm{s}}
\newcommand{\brhos}{\bar\rho_\mathrm{s}}
\newcommand{\rhooff}{\rho_\mathrm{conf}} 
\newcommand{\rhoref}{\rho_\mathrm{ref}}
\newcommand{\rhoHxc}{\rho_\mathrm{Hxc}}
\newcommand{\rhob}{\rho_\mathrm{b}}
\newcommand{\rhof}{\rho_\mathrm{f}}
\newcommand{\jjs}{\bm{j}_\mathrm{s}}
\newcommand{\jjMregHxc}{\bjj_{\mathrm{Hxc}}}

\newcommand{\AAs}{\bm{A}_\mathrm{s}}
\newcommand{\bAAs}{\bAA_\mathrm{s}}
\newcommand{\AAref}{\bm{A}_\mathrm{ref}}

\newcommand{\potSpace}{\mathcal{P}}
\newcommand{\potSpaceVec}{\mathcal{P}_\mathrm{vec}}
\newcommand{\densSpace}{\mathcal{P}^*}
\newcommand{\densSpaceVec}{\mathcal{P}_\mathrm{vec}^*}
\newcommand{\X}{\mathcal{X}}

\newcommand{\mf}{\mathrm{mf}}
\newcommand{\Eelm}{\mathcal{E}_\mathrm{elm}}

\begin{document}

\title{Quantum electrodynamics of equilibrium systems:\\A rigorous Maxwell-regularized functional-theoretic formulation}

\author{Markus Penz}
\email[Electronic address:\;]{m.penz@inter.at}
\affiliation{Arnold Sommerfeld Center for Theoretical Physics,
\mbox{Ludwig-Maximilians-Universit\"at M\"unchen, Germany}}
\affiliation{Department of Computer Science, Oslo Metropolitan University,  Norway}
\affiliation{\mbox{Max Planck Institute for the Structure and Dynamics of Matter and Center for Free-Electron Laser Science, Hamburg, Germany}}
\author{Christian Jöns}
\affiliation{\mbox{Max Planck Institute for the Structure and Dynamics of Matter and Center for Free-Electron Laser Science, Hamburg, Germany}}
\author{Michael Ruggenthaler}
\email[Electronic address:\;]{michael.ruggenthaler@mpsd.mpg.de}
\affiliation{\mbox{Max Planck Institute for the Structure and Dynamics of Matter and Center for Free-Electron Laser Science, Hamburg, Germany}}
\author{Angel Rubio}
\affiliation{\mbox{Max Planck Institute for the Structure and Dynamics of Matter and Center for Free-Electron Laser Science, Hamburg, Germany}}
\affiliation{Initiative for Computational Catalysis, The Flatiron Institute, Simons Foundation, \mbox{New York City, USA}}

\begin{abstract}
A framework for quantum electrodynamics of equilibrium systems is proposed that takes the Maxwell equations as foundational and links them to the structure of density-functional theory for the quantum system.
By switching from purely internal observables like the one-particle density to combined internal and external quantities, the external sources and their energy are taken into account. A fully regularized functional theory emerges that avoids the usual representability problems.
With the Kohn--Sham construction that links to an auxiliary uncoupled system, the ultra-violet cutoff can be removed and one arrives at a fully renormalized and non-perturbative light-matter description. Regularized forms of density-functional theory with and without magnetic fields appear as boundary cases, while the emergent physical picture reproduces the macroscopic Maxwell equations.
\end{abstract}

\maketitle
\tableofcontents

\section{Introduction}
\label{sec:intro}

Light-matter interactions lie at the heart of modern quantum physics. For instance, the ubiquitous Schr\"odinger equation can be derived from \ac{QED} in the Coulomb gauge, assuming small (non-relativistic) momenta for the charged particles. Specifically, one realizes that the Coulomb interaction arises due to quantizing matter \emph{and} light at the same time and that it describes the effects of the longitudinal electromagnetic field energy~\cite{greiner2013field}. Thus the Coulomb interaction is \emph{not} a property of matter alone. Overall, \ac{QED} highlights that the distinction between light and matter is ambiguous. This is visible, on the one hand, from the fact that the basic Hamiltonians (and with them the wave functions) allow for gauge transformations that mix light and matter degrees of freedom. For example, in Coulomb gauge, the energy of the longitudinal electromagnetic field is described by a matter operator, the usual Coulomb interaction. On the other hand, the basic building blocks of our light-matter quantum theories are unobservable \emph{bare} objects, i.e., bare charged particles and bare photons, which are just theoretical tools to set up the physically relevant \emph{interacting} theory that then leads to \emph{dressed} objects with observable properties. For instance, the masses of the \emph{observable} charged particles are renormalized and contain bound photons arising from self-interactions~\cite{greiner2013field,spohn-book,ruggenthaler2023understanding}. In other words, even a single free electron contains already photonic contributions. Moreover, we learn from \ac{QED} that we need to set an \emph{external} scale at which we consider our interacting theories. This is to say that external classical parameters (source terms or fields) are needed to extract physical information from an interacting theory~\cite{Peskin:1995ev, Ryder:1996fk, Haag1996LocalQuantumPhysics}.

While these facts are well-established in high-energy physics, they are often overlooked in the context of condensed-matter physics and quantum chemistry~\cite{ruggenthaler2023understanding}. That there is a mismatch between the different fields is largely due to the fact that full \ac{QED} is a perturbative theory for scattering events, while most questions in condensed-matter physics and quantum chemistry concern non-perturbative bound-state problems. So fundamentally, these fields use different mathematical tools and only under certain approximations can they be connected. The main issue lies in the fact that \ac{QED} needs to be regularized to be applicable beyond perturbation theory, which in its simplest form means to introduce smallest (ultra-violet) and largest (infrared) length and time scales through cutoffs. In free space, which obeys the full Poincar\'e symmetry and which is the standard setting of \ac{QED}, such scales are not given a priori and hence tremendous effort has been put into making the theory independent of them, i.e., to find a fully non-perturbatively renormalized \ac{QED} theory without cutoffs.

In low-energy physics, however, such scales are usually already part of the initial formulation. For instance, in order to define a molecular structure, the Born--Oppenheimer approximation is commonly invoked, which perfectly localizes the nuclei, breaks the symmetry of free space, and introduces reference scales~\cite{Sutcliffe2012}. This way, symmetry-breaking and scales are introduced by external confining potentials. In such a setting, it can be shown rigorously that within non-relativistic \ac{QED} a bound state exists and that the infrared cutoff can be removed non-perturbatively~\cite[Sec.~15.1]{spohn-book}. Whether also the ultra-violet regularization can be removed is an open question in this wave-function formulation. If this can be achieved, we would have a fully renormalized non-perturbative theory of coupled light and matter in the low-momentum (for individual bare matter particles) approximation, which would be a tremendous step forward. However, other problems still remain, like the fact that the theory is not explicitly Lorentz covariant. Moreover, for all practical purposes, the theory is still computationally infeasible on the wave-function level due to the high dimensionality of the underlying Hilbert space that has been introduced for a linear many-body description.

This computational infeasibility, which similarly holds for anything but the simplest few-particle systems in \ac{QM}, has led to the development of \ac{DFT}~\cite{parr,eschrig2003-book,vonBarth2004basic,burke2007abc,dreizler-gross-book}. It is a \emph{formally exact} reformulation of \emph{linear} \ac{QM} in terms of a \emph{non-linear} theory for the single-particle (probability or charge) density. This reformulation makes many-body \ac{QM} computationally tractable, albeit for the price of a dependence on approximations to the unknown exchange-correlation energy~\cite{dreizler-gross-book,EngelDreizler2011,Toulouse2022-chapter,tran2026} or force~\cite{MarquesMaitraNogueiraGrossRubio2012,Ullrich2011,tancogne2024exchange}. Yet, even simple approximations lead to reasonable predictions, and hence \ac{DFT} and its variants have become the most widely used electronic-structure method in physics and chemistry~\cite{Jones2015}. Besides its practical relevance, \ac{DFT} provides an alternative view on the basics of \ac{QM}: It shows that the gauge-dependent and unobservable wave function can be replaced exactly by a physical and gauge-independent quantity when evaluating ground-state properties. Despite this success, there are still some important open questions in the foundations of \ac{DFT}~\cite{Lammert2007,Teale2022-DFTexchange,Penz2023-PartI}, which have hitherto only been resolved rigorously for special cases~\cite{Sutter2024,CarvalhoCorso2025}, or by supplemental regularization of the theory. A special transformation that overcomes most theoretical problems is Moreau--Yosida regularization of the universal functional of \ac{DFT}~\cite{Kvaal2014,KSpaper2018,kvaal2022-chapter,MY-Perspective}. This regularization is \emph{lossless} in the sense that it makes the same predictions as standard \ac{DFT}, but the regularized functional and its arguments loose their physical relevance, which also means that to-date this method has limited practicability (with the notable exception of its application in Kohn--Sham inversion algorithms~\cite{Penz2023-MY-ZMP,Herbst2025,Bohle2026,MY-periodic}).

In this work, we will overcome most of the above mentioned issues by going back to the basics of light-matter interactions and by starting from the Maxwell equations as the fundamental theory. Instead of focusing on the quantum system alone, we include the external charge density and current into the basic descriptors of the theory and require that their electromagnetic field has finite energy. This combination into mixed quantities is possible due to an exact mathematical duality between electromagnetic potentials and matter quantities mediated through the Maxwell equations. Carried over to the universal functional this then automatically leads to Moreau--Yosida regularization of the Pauli--Fierz theory and allows for a rigorous reformulation of \emph{linear} non-relativistic equilibrium \ac{QED} as a \emph{non-linear} functional theory based solely on physical observables. We suggest calling this new theory \ac{MaxQEDFT}. By applying the Kohn--Sham construction~\cite{kohn1965self} to this theory, we can show that the ultra-violet cutoff of non-relativistic \ac{QED} can be removed non-perturbatively. As a straightforward consequence, we show how we can also overcome the mathematical issues of standard \ac{DFT} and how to arrive at a mean-field formulation that includes magnetic fields.

\section{Setting}
\label{sec:setting}

In this work, we are interested in equilibrium states of condensed-matter systems (atoms, molecules, solids). Yet, from the start, we encounter a fundamental issue in the theoretical formulation of this problem. Since the free-space theories on $\mathbb{R}^3$ are \emph{too} symmetric, specifically, since they include translational and rotational invariance for the \emph{total} system, they only provide scattering (non-stationary) solutions. Thus, without further modifications of the theory, we cannot expect equilibrium solutions. This well-known issue leads to conceptual problems in quantum chemistry and solid-state physics~\cite{Sutcliffe2012} as well as in quantum thermodynamics~\cite{thirring2002quantum}. The common solution is to break the (too high) symmetry of the free-space theories and localize the system such that ground-state solutions become possible. This can be achieved, for instance, by considering the system in a finite (thermodynamically extensive) volume or by enclosing it in a confining potential. Both cases imply that a system of charged quantum particles is in contact with an electromagnetic environment.

In this spirit, we begin with the static Maxwell equations (in four-potential form with Coulomb gauge, as used throughout this work)
\begin{align}
-\Delta \phi &= \frac{\sigma}{\epsilon_0},\label{eq:Gauss}
\\
\nabla \times \nabla \times \aa  &= \mu_0 \jj, \label{eq:AmpereMaxwell}
\end{align}
where $\epsilon_0$ and $\mu_0$ are the vacuum permittivity and permeability, respectively. The $\sigma$ and $\jj$ are the charge density and (transverse) current of either the equilibrium system or the environment, and both need to obey the static continuity equation
\begin{align}
- \nabla \cdot \jj = \partial_t\sigma = 0. \label{eq:chargeconservation}
\end{align}
We use $\sigma$ and $\aa$ here for generic charge densities and vector potentials instead of the more usual $\rho$ and $\AA$ that will have a special meaning as the expectation values of the quantum system later.
Since we are in Coulomb gauge, which is a complete gauge, the scalar and vector potentials are uniquely determined and directly represent the respective physical fields
\begin{align}
\EE = -\nabla \phi, 
\quad
\BB = \nabla \times \aa. \label{eq:magneticfield}
\end{align}

In our description, a light-matter quantum system with internal density $\rho$ and vector potential $\AA$ is complemented by and coupled to a classical environment governed by the Maxwell equations. It is described by an electrostatic potential $\phi$ and a charge current $\jj$ that together provide the means to steer the quantum system.
This reproduces the setting of standard \ac{QED} and \ac{QEDFT}~\cite{ruggenthaler2017-QEDFT,ruggenthaler2023understanding, ruggenthaler2014quantum}.
But here the different external conditions should not simply be expressed by fixed external parameters, but receive their own physical reality and energetic description.
The quantum system can then also act back on the environment which affords a symmetric description of system and environment. This is similar to the geometry optimization in the Born--Oppenheimer picture~\cite{Schlegel2011,Jecko2014} or polarizable-medium models~\cite{Tomasi2005,Mennucci2012,Herbert2021}.
Including the energy content of the external sources into the overall description will have surprising consequences.

We still need to decide on a confining potential that localizes our system in the lab frame and will serves as an offset for all possible potentials that add to it. Since the details of the equilibrium system will not depend on the precise form of confinement~\cite{thirring2002quantum,spohn-book} (as long as we can make the potential arbitrarily shallow in the form of a free-space limit), we choose a harmonic potential. This choice is not arbitrary but has a clear meaning in the context of coupled light-matter systems. Using Eq.~\eqref{eq:Gauss}, we find that a harmonic confining potential for negatively charged particles,
\begin{align}
\phiconf(\rr) = -\frac{1}{2}\frac{\gamma}{\epsilon_0} \rr^2 \leq 0,
\end{align}
implies a constant positive charge density,
\begin{align}
-\epsilon_0\Delta \phiconf = \rhooff \quad\Rightarrow\quad \rhooff(\rr) =  3 \gamma \geq 0.
\end{align}
This charge density $\rhooff$ can be directly connected to a homogeneous positive charge background
$\rhooff(\rr) = n_{\rm conf} e$,
where $n_{\rm conf}$ is the number of positive charges per unit volume and $e>0$ is the (positive) elementary charge. This is the free-space analogue~\cite{Loos2012} of the periodic-boundary jellium model~\cite{brack1993physics}. Indeed, due to Kohn's theorem~\cite{Kohn1961}, we find that a many-electron system in a harmonic potential of the form 
\begin{align}
\phiconf(\rr) = - \frac{1}{2} \frac{n_{\rm conf}e}{3 \epsilon_0} \rr^2 = -\frac{1}{2} \frac{m}{3e} \omega_{\rm p}^2 \rr^2,
\end{align}
with $m$ the \emph{observable mass} of the electrons, and with a plasma frequency
\begin{align}\label{eq:Mie_frequency}
\omega_{\rm p}^2 = \frac{n_{\rm conf} e^2}{\epsilon_0 m},
\end{align}
displays the expected Mie-plasmon oscillations $\omega_{\rm Mie} = \omega_{\rm p}/\sqrt{3}$ of a charge-neutral isotropic system.
At the same time, the confined electrons lead to a natural infrared regularization for the Maxwell field. For $\omega > \omega_{\rm p}$ electromagnetic fields propagate freely within the neutral plasma, while for $\omega< \omega_{\rm p}$ electromagnetic modes will be suppressed. One can understand this from the transverse dispersion relation~\cite{jackson-book}
\begin{align}
k^2 c^2 = \omega^2 \epsilon(\omega) = \omega^2\left(1-\frac{\omega_{\rm p}^2}{\omega^2}\right), 
\end{align}
which for $\omega < \omega_{\rm p}$ allows only imaginary solutions. 

As stressed before, the confining potential $\phiconf$ has the important role to break the symmetry of free space, to initially localize the quantum system, and to guarantee existence of a ground state. It also ensures that there is a reference system that fixes the charge sector of the system. How all of this appears in non-relativistic \ac{QED} will be discussed in Sec.~\ref{sec:qed-and-reference}.
We add that the origin of the confining potential does not necessarily need to be of electrostatic nature but could also be gravitational. In our current setting, we do not include a weakly coupled heat bath to our setting. However, since we consider a quantized photon field, the matter subsystem alone will never be at (quantum-mechanical) zero temperature. That is, upon tracing out the coupled photonic system, the matter subsystem is not in a pure state but resembles a thermal state~\cite{sidler2023numerically}.
Before we now turn to the description of the light-matter quantum system, we need to formalize the interaction between system and electromagnetic environment, where the Maxwell equations link potentials in a unique way to charge densities and currents. The respective problem-adapted Hilbert spaces for potentials on one side and charge densities and currents on the other will become central to the whole theory, allowing to treat the environment and the internal expectation values of the quantum system in a unified manner.

\section{Maxwell equations in Hilbert spaces}
\label{sec:Maxwellhilbert}

As a first step in formalizing our physical setup, we put the static Maxwell Eqs.~\eqref{eq:Gauss}-\eqref{eq:AmpereMaxwell} into a Hilbert-space form that is compatible with quantum physics. We ask the reader to bear with us, as this section deals with mathematical technicalities, but we will introduce some important notation here that we heavily rely on during the later sections. Since self-adjoint operators are equivalent to the inner product of problem-adapted Hilbert spaces~\cite[Th.~3.17]{lewin2024-book}, we are able to encode the whole Maxwell equations into the topology of such spaces. This, at the same time, allows a canonical mapping from elements of this Hilbert space to uniquely defined topologically dual elements that represent solutions of the Maxwell equations.
In flat space-time and using the vector-Laplacian identity $\Delta_\mathrm{vec}=\nabla\nabla\cdot - \nabla\times\nabla\times$ together with Coulomb gauge ($\nabla\cdot\aa=0$), the two Eqs.~\eqref{eq:Gauss} and \eqref{eq:AmpereMaxwell} take the form of component-wise vacuum Poisson equations
\begin{align}
    \label{eq:Poisson-phi}
    &-\Delta \phi = \frac{\sigma}{\epsilon_0}, \\
    \label{eq:Poisson-a}
    &-\Delta a_k = \mu_0 j_k, \quad k\in\{1,2,3\}.
\end{align}
Those quantities bear the usual SI units, with potential $[\phi]=\mathrm{V}$, charge density $[\sigma]=\mathrm{C}/\mathrm{m}^3$, vector potential $[a_k]=\mathrm{Vs}/\mathrm{m}$, and charge-current density $[j_k]=\mathrm{A}/\mathrm{m}^2$.
The function spaces for electromagnetic potentials that already embed these Poisson equations geometrically (we will see below what this means) are defined as
\begin{align}
    &\potSpace = \{ \phi \in \Lloc(\R^3) \big/ \R \mid
    \nabla \phi \in (L^2(\R^3))^3 \}, \\
    &\potSpaceVec = \{ \aa=(a_1,a_2,a_3) \mid a_k \in \potSpace, \nabla \cdot \aa = 0 \}.
\end{align}
This just means that each field component is a real integrable function on $\R^3$, but already modulo constants as customary for potentials. The weakly-defined gradients of each component are all assumed square-integrable and we further enforce Coulomb gauge for vector potentials $\aa\in\potSpaceVec$. We first limit the discussion to the space $\potSpace$ whose properties carry over to its vector version $\potSpaceVec$. Together with the inner product
\begin{equation}\label{eq:inner-prod}
    \langle \phi,\phi' \rangle_\potSpace = \epsilon_0 \int \nabla \phi(\rr) \cdot \nabla \phi'(\rr) \, \rmd\rr
\end{equation}
and its induced norm
\begin{equation}\label{eq:pot-norm}
    \|\phi\|_\potSpace = \left( \epsilon_0 \int |\nabla \phi(\rr)|^2 \, \rmd\rr \right)^{1/2} = \sqrt{\epsilon_0} \|\nabla \phi\|_{L^2}
\end{equation}
the space $\potSpace$ is a Hilbert space, called a \emph{homogeneous Sobolev space}~\cite{mazya-book,galdi-book,ortner2012}. This norm has already found applications in physics-informed regularization within numerical methods in quantum chemistry~\cite{Heaton-Burgess2007,Medvidovic2026}.
Here and in the following $\phi\in\potSpace$ usually means to take any representative from the equivalence classes in $\potSpace$. Having the potential defined only up to an additive constant is indeed necessary such that the norm from Eq.~\eqref{eq:pot-norm} is positive definite.
We further note the continuous embedding into Lebesgue-integrable spaces~\cite[Th.~2.2(i)]{ortner2012}
\begin{equation}\label{eq:hom-Sobolev-embedding}
    \potSpace \subseteq L^{6}(\R^3).
\end{equation}
This means there is an $M>0$ such that for every $\phi\in\potSpace$ we find a $c\in\R$ and have $\phi-c\in L^6$, which reflects the definition modulo a constant, and $\|\phi-c\|_{L^6}\leq  M\|\phi\|_\potSpace$. This embedding will become important later and from here on we suppress the domain $\R^3$ from the notation of $L^p$ spaces.

Any Hilbert space is isomorphic to its topological dual space $\potSpace^*$ that is the set of bounded linear functionals from $\potSpace$ to $\R$ and forms a Hilbert space itself. A precise characterization of the spaces $\potSpace$ and $\potSpace^*$ will be given in a separate mathematical publication, but we need to comment on a few important properties for later use. There exists a canonical mapping $\potSpace^*\to\potSpace$ from the Riesz representation theorem that is linear, isometric, and bijective and assigns every $\sigma\in \potSpace^*$ an element $\phi\in\potSpace$ with $\|\phi\|_\potSpace^2=\|\sigma\|_{\potSpace^*}^2=\sigma(\phi)$, such that for all $\phi'\in\potSpace$
\begin{equation}\label{eq:P-dual-pairing}
    \sigma(\phi') = \langle \sigma,\phi' \rangle = \langle \phi,\phi' \rangle_\potSpace.
\end{equation}
Here, $\langle \sigma,\phi' \rangle$ (written without the index $\potSpace$ for the space) denotes the dual pairing on $\potSpace^*\times\potSpace$ and is just a way how to rewrite $\sigma(\phi')$ in the case of linear functionals. We will conveniently denote the important Riesz mapping, also called the duality mapping, as $\sigma\mapsto \sigma^*$ (so just with a star) and equally its inverse $\phi\mapsto \phi^*$ that maps $\potSpace\to\densSpace$. Note that it is linear, $(\alpha \phi+\beta \phi')^*=\alpha \phi^*+\beta (\phi')^*$ for all $\phi,\phi'\in \potSpace$, $\alpha,\beta\in \R$, and bijective, so $\phi=\sigma^*$ if and only if $\sigma=\phi^*$ and thus $\phi^{**}=\phi$ and likewise $\sigma^{**}=\sigma$.
From this definition, we also have $\|\phi\|_\potSpace^2 = \langle \phi^*,\phi \rangle$, $\|\sigma\|_{\densSpace}^2 = \langle \sigma,\sigma^* \rangle$, and $\langle \sigma,\phi \rangle = \langle \phi^*,\sigma^* \rangle$.
Using the explicit form of Eq.~\eqref{eq:inner-prod} with $\sigma = \phi^*$ means that after partial integration, it holds for all test functions $\varphi$ that
\begin{equation}
    \langle \sigma,\varphi \rangle = -\epsilon_0 \int \varphi(\rr) \Delta \phi(\rr)\, \rmd\rr.
\end{equation}
But since the test functions are dense in $\potSpace$~\cite{ortner2012}, this tells us that $\phi=\sigma^*$ is exactly the unique solution of the (weak) Poisson equation from Eq.~\eqref{eq:Poisson-phi} for a density $\sigma \in \potSpace^*$, which means it holds
\begin{equation}\label{eq:Poisson}
    \sigma=\phi^* = -\epsilon_0\Delta\phi.
\end{equation}
We can thus identify the dual space $\densSpace$ of the potential space $\potSpace$ as the space of densities.
Such a density $\sigma=\phi^*$, since it is a linear functional on $\potSpace$, might also be distributional~\cite[Sec.~3.1]{ortner2012}.

Exactly the same properties now hold for the vector-potential space $\potSpaceVec$ and its dual $\densSpaceVec$ that correspondingly describes density currents. The only differences are that the vacuum permittivity $\epsilon_0$ gets replaced with the reciprocal vacuum permeability $\mu_0^{-1}$ in all equations and that all $\jj\in\densSpaceVec$ also have
$\nabla \cdot \jj = 0$ (zero divergence)
as a result of the Coulomb gauge for elements in $\potSpaceVec$.
That is, the currents are transverse and obey local charge conservation according to Eq.~\eqref{eq:chargeconservation}. The corresponding canonical mapping $\potSpaceVec\to\potSpaceVec^*$ is then
\begin{equation}
        \jj = \aa^* = -\mu_0^{-1}\Delta_\mathrm{vec} \aa = \mu_0^{-1}\nabla\times\nabla\times\aa.
\end{equation}

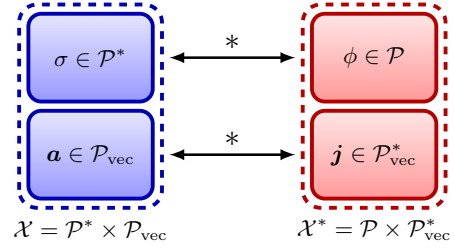
\begin{figure}
\resizebox{.7\columnwidth}{!}{%
\begin{tikzpicture}[
  box/.style={
    rectangle, rounded corners=5pt,
    draw=blue!70!black, line width=1.5pt,
    top color=blue!10, bottom color=blue!40,
    minimum width=1.7cm, minimum height=1.2cm,
    inner sep=5pt, align=center,
  },
  boxalt/.style={
    rectangle, rounded corners=5pt,
    draw=red!70!black, line width=1.5pt,
    top color=red!10, bottom color=red!40,
    minimum width=1.7cm, minimum height=1.2cm,
    inner sep=5pt, align=center,
  }
]
\node[box] (boxrho) {$\sigma\in\densSpace$};
\node[box] (boxA) at (boxrho.south) [yshift=-0.7cm] {$\aa\in\potSpaceVec$};
\draw[dashed, draw=blue!70!black, line width=1.5pt, rounded corners=7pt] 
    ([xshift=-1mm,yshift=-1mm]boxA.south west) 
    rectangle 
    ([xshift=1mm,yshift=1mm]boxrho.north east);
\node at (boxA.south) [yshift=-0.4cm] {$\X = \densSpace\times\potSpaceVec$};
\node[boxalt] (boxphi) at (boxrho.east) [xshift=3cm] {$\phi\in\potSpace$};
\node[boxalt] (boxj) at (boxphi.south) [yshift=-0.7cm] {$\jj\in\densSpaceVec$};
\draw[dashed, draw=red!70!black, line width=1.5pt, rounded corners=7pt] 
    ([xshift=-1mm,yshift=-1mm]boxj.south west) 
    rectangle 
    ([xshift=1mm,yshift=1mm]boxphi.north east);
\node at (boxj.south) [yshift=-0.4cm] {$\X^* = \potSpace\times\densSpaceVec$};
\draw[latex-latex,line width=1pt] ([xshift=.2cm]boxrho.east) -- node[midway,above] {\large$*$} ([xshift=-.2cm]boxphi.west);
\draw[latex-latex,line width=1pt] ([xshift=.2cm]boxA.east) -- node[midway,above] {\large$*$} ([xshift=-.2cm]boxj.west);
\end{tikzpicture}
}
\caption{Schematics for the principal Hilbert spaces $\X$ for densities and vector potentials as the internal quantities (later evaluated as expectation values on the quantum system) and its dual $\X^*$ for the external sources, linked through Maxwell duality.}
\label{fig:spaces}
\end{figure}

We are now in the position to combine those spaces into the four-vector Hilbert space $\potSpace\times\potSpaceVec$ and its dual $\densSpace\times\densSpaceVec$.
The norm squared of a four-potential $(\phi,\aa)\in\potSpace\times\potSpaceVec$ then gives
\begin{equation}\label{eq:field-energy}
\begin{aligned}
    \frac{1}{2}&\|(\phi,\aa)\|_{\potSpace\times\potSpaceVec}^2 = \frac{1}{2}\langle\phi^*,\phi\rangle+\frac{1}{2}\langle\aa^*,\aa\rangle\\
    &= -\frac{\epsilon_0}{2}\int\phi(\rr)\Delta\phi(\rr)\rmd\rr - \frac{1}{2\mu_0} \int \aa(\rr) \cdot \Delta_\mathrm{vec} \aa(\rr) \, \rmd\rr \\
    &= \frac{\epsilon_0}{2}\|\nabla\phi\|_{L^2}^2 + \frac{1}{2\mu_0} \|\nabla \times \aa\|_{L^2}^2.
\end{aligned}
\end{equation}
We immediately notice the relevance of this norm for the discussion of electromagnetism, since this gives exactly the energy of the electromagnetic field. In other words, any potential in $\potSpace\times\potSpaceVec$ has finite electromagnetic energy and the respective norm can aptly be called an energy norm.
We also want to study the norm of the dual space, where we can write the solution for the Poisson Eq.~\eqref{eq:Poisson} in terms of a Green's function (Riesz potential) if $\sigma$ and the components $j_k$ are integrable. We then have
\begin{align}
    &\phi(\rr) = \sigma^*(\rr) = \int \frac{\sigma(\rr')}{4\pi\epsilon_0|\rr-\rr'|}\,\rmd\rr', \label{eq:f-star}\\
    &a_k(\rr) = j_k^*(\rr) = \int \frac{\mu_0 j_k(\rr')}{4\pi|\rr-\rr'|}\,\rmd\rr'.
\end{align}
Note how this corresponds to the Hartree potential of electronic-structure theory. This way, $(\phi,\aa)=(\sigma^*,\jj^*)=(\sigma,\jj)^*$ are exactly the solutions to the static Maxwell equations with the sources $(\sigma,\jj)$, which prompted us to call the duality structure of this Hilbert space the \emph{Maxwell duality}.
This further allows us to rewrite the norm squared in $\densSpace\times\densSpaceVec$ as
\begin{equation}\label{eq:HartreeQED}\begin{aligned}
    &\|(\sigma,\jj)\|_{\densSpace\times\densSpaceVec}^2 = \langle(\sigma,\jj),(\sigma,\jj)^*\rangle = \langle \sigma,\sigma^* \rangle + \langle \jj,\jj^* \rangle \\
    &= \iint \frac{\sigma(\rr)\sigma(\rr')}{4\pi\epsilon_0|\rr-\rr'|} \, \rmd\rr\rmd\rr' + \iint \frac{\mu_0 \jj(\rr)\cdot\jj(\rr')}{4\pi|\rr-\rr'|} \, \rmd\rr\rmd\rr',
\end{aligned}\end{equation}
which corresponds to (twice) the classical Hartree energy of the charge density and current. Although those four-vector spaces are very elegant, they will not constitute the central spaces for our theory in this form. Instead, we group the density of the quantum system with the expectation value of the vector-potential operator, which together represents the reduced quantities for the quantum system. In terms of spaces, this means we take $\densSpace\times\potSpaceVec$ for the internal quantities and its dual space $\potSpace\times\densSpaceVec$ to describe the classical scalar potential (coupling to the charged quantum particles) and charge currents (coupling to the photon degrees-of-freedom). So, for later use, we define our basic spaces
\begin{align}
    \X = \densSpace\times\potSpaceVec \quad\text{and}\quad
    \X^* = \potSpace\times\densSpaceVec
\end{align}
that are also illustrated in Fig.~\ref{fig:spaces}. Their norms are given by
\begin{align}
    \label{eq:norm-X}
    &\|(\sigma,\aa)\|^2_{\X} = \|\sigma\|_{\densSpace}^2 + \|\aa\|_{\potSpaceVec}^2,\\
    \label{eq:norm-Xstar}
    &\|(\phi,\jj)\|^2_{\X^*} = \|\phi\|_{\potSpace}^2 + \|\jj\|_{\densSpaceVec}^2.
\end{align}
These norms describe, again, (twice) the intrinsic energy of the electromagnetic field when $\phi=\sigma^*$ and $\aa=\jj^*$,
\begin{equation}\label{eq:def-F-elm}
    \Eelm(\phi,\jj) = \frac{1}{2}\|(\phi,\jj)\|^2_{\X^*}.
\end{equation}
The reason for pairing a potential $\phi\in\potSpace$ with a current $\jj\in\densSpaceVec$ here is that these will be the parameters in the Pauli--Fierz Hamiltoinian and thus appear as external variables in \ac{QEDFT}.
For later purpose, we calculate the functional derivatives of the norm-squares like in Eqs.~\eqref{eq:norm-X} and \eqref{eq:norm-Xstar}, which just gives the duality mapping,
\begin{align}\label{eq:square-deriv-X}
    &\rmd\! \left((\sigma,\aa)\mapsto\frac{1}{2}\|(\sigma,\aa)\|^2_{\X}\right) = (\sigma,\aa)^*\in\X^*,\\
    &\rmd\! \left((\phi,\jj)\mapsto\frac{1}{2}\|(\phi,\jj)\|^2_{\X^*}\right) = \rmd\Eelm(\phi,\jj)= (\phi,\jj)^*\in\X.
\end{align}

The setting that we constructed here around the Maxwell equations is based on Hilbert spaces. This does not only remind of \ac{QM}, it is also the starting point to derive the Riemann--Silberstein equation, and with it the Maxwell equations, from first principles~\cite{ruggenthaler2023understanding}. By choosing the Coulomb gauge and second quantization of the transverse part of the electromagnetic four-potential, one then arrives at the field-part of the \ac{QED} Hamiltonian with quantized vector potential $\hAA$.
The zero-component of the four-potential, on the other hand, enters a Poisson equation and gets quantized via the matter degrees-of-freedom which yields the Coulomb interaction. This explains why the space $\X$ for the quantum system that gathers all internal quantities is composed from a space for \emph{densities} and \emph{transverse vector potentials}.
Via minimal coupling of $\hAA$ to the Schr\"{o}dinger equation this procedure then yields the fundamental Hamiltonian of non-relativistic \ac{QED}, the Pauli--Fierz Hamiltonian, which will be the topic of the next section.

\section{QED Hamiltonian and reference system}
\label{sec:qed-and-reference}

As a next step, we discuss the class of non-relativistic \ac{QED} Hamiltonians that will allow a rigorous functional-theoretic reformulation for coupled light-matter systems. Mathematical details on function spaces, operator domains and existence as well as uniqueness theorems can be found in the excellent book of \citet[Sec.~13.3 and 13.4]{spohn-book}. To formulate a well-defined \ac{QED} theory, one needs to introduce a `form factor' smearing out the point-like character of the light-matter interaction and thus regularizing on the small length (high frequency) scale. One does this already in the Fourier domain with a function $\widehat{\varphi}^\Lambda(\kk)$ that is symmetric and cuts off the high-frequency contributions. Many choices are possible, but the simplest one is the characteristic function
\begin{equation}\label{eq:form-factor}
    \widehat{\varphi}^\Lambda(\kk) = \left\{ \begin{array}{ll}
        (2\pi)^{-3/2} & \text{for}\; |\kk| < \Lambda \; \text{and} \\
        0 & \text{else},
    \end{array} \right.
\end{equation}
where $\Lambda$ is the ultra-violet cutoff. We then have $\varphi^\Lambda(\rr)\to\delta(\rr)$ as $\Lambda\to\infty$ and the cutoff gets removed. The introduction of the form factor influences various parts of the Hamiltonian, and we will further denote the dependence on it by $\Lambda$. After second-quantization of the free-space field modes including the form factor, the transverse vector-potential operator has the form
\begin{equation}\label{eq:A-def}
\begin{aligned}
    \hAA^\Lambda(\rr) = \sqrt{\frac{\hbar}{\epsilon_0}}\sum_{\lambda =1}^{2}\int & \frac{\be(\kk,\lambda)\widehat{\varphi}^\Lambda(\kk)}{\sqrt{2 \omega_{\kk}}} \Big( \hat{a}(\kk,\lambda) \rme^{\rmi \kk \cdot \rr} \\
	& + \hat{a}^{\dagger}(\kk,\lambda) \rme^{- \rmi \kk \cdot \rr} \Big) \rmd \kk.
\end{aligned}
\end{equation}
Here, $\omega_{\kk} = c|\kk|$ and the creation and annihilation operators for the field modes obey the usual canonical commutation relations~\cite[Sec.~13.2]{spohn-book}.
The polarization directions are chosen to fulfill
\begin{equation}
\kk\cdot\be(\kk,\lambda) = 0,\; \be(\kk,1) \cdot \be(\kk,2) = 0,\; |\be(\kk,\lambda)| = 1.
\end{equation}
Note that while this choice for $\be(\kk,\lambda)$ is by far the most common, other constructions that might be advantageous due to better continuity properties are possible~\cite{Lieb2004}.

In the limit $\Lambda\to 0$ we have $\hAA^\Lambda(\rr) \to 0$, whereas removing the cutoff $\Lambda\to\infty$ in the coupling to the quantized electrons does \emph{not} retain a well-defined operator. This is an expression of the renormalization problem of \ac{QED}. An alternative method to remove the cutoff within the formalism of \ac{QEDFT} will be discussed in Sec.~\ref{sec:renorm-ks}.
As mentioned in Sec.~\ref{sec:intro}, the zero-component of the four-potential is quantized via the matter part and yields the regularized Coulomb interaction
\begin{equation}\label{eq:interact-limit-Coulomb}
    \wC^\Lambda(\rr) = \int \frac{|\widehat{\varphi}^\Lambda(\kk)|^2}{\epsilon_0 |\kk|^2}\rme^{-\rmi\kk\cdot\rr} \rmd\kk \left\{ \begin{array}{cc}
        \overset{\Lambda\to\infty}{\longrightarrow} & 1/(4 \pi \epsilon_0 |\rr|) \\
        \overset{\Lambda\to 0}{\longrightarrow} & 0.
    \end{array} \right.
\end{equation}
We could directly work with the full Coulomb interaction ($\Lambda \rightarrow \infty$), as it is common practice in \ac{QM} and non-relativistic \ac{QED}, since no normalization problem appears here. Keeping the regularization also on the longitudinal interaction will, however, clarify the connection between the Kohn--Sham construction and renormalization theory, as discussed in Sec.~\ref{sec:renorm-ks}.

The Pauli--Fierz Hamiltonian~\cite[Sec.~13.2]{spohn-book} for $N$ electrons with particle positions $\rr_i$ coupling to the photon field and given external potential and charge current $(\phi,\jj)\in\X^*$ (taken from the space introduced in Sec.~\ref{sec:Maxwellhilbert}) is then
\begin{equation}\label{eq:H}
\begin{aligned}
\hat{H}^\Lambda_{\phi,\jj} &= \sum_{i=1}^{N} \frac{1}{2m^\Lambda}\left[\bm{\sigma}_i\cdot \left(-\rmi\hbar\nabla_i + e\hAA^\Lambda(\rr_i) \right)  \right]^2 \\
& + \sum_{i<j}^{N} e^2\wC^\Lambda(\rr_i-\rr_j) + \hat H_\mathrm{field}
\\
& + \int \hat\rho(\rr)(\phiconf(\rr) + \phi(\rr))\rmd \rr - \int\hAA^\Lambda(\rr) \cdot \jj(\rr)\rmd \rr. 
\end{aligned}
\end{equation}
Here, $m^{\Lambda} > 0$ is the \emph{bare} electronic mass that depends on the chosen cutoff and the field-only part of the Hamiltonian is
\begin{equation}\label{eq:H-field}
    \hat H_\mathrm{field} = \sum_{\lambda=1}^2 \int_{|\kk|<\Lambda} \hbar\omega_{\kk}\hat{a}^{\dagger}(\kk,\lambda)\hat{a}(\kk,\lambda) \rmd\kk.
\end{equation}
In the coupling to the external potential, the charge-density operator for $N$ electrons is given by $\hat{\rho}(\rr) = -e \sum_{i=1}^{N} \delta(\rr-\rr_i)$. Note that this operator-valued distribution does not only give units of charge per volume in its expectation value, but also bears the negative charge of electrons. This is necessary for the expectation value to correctly connect to the charge distribution that occurs in the Maxwell equations.
The different signs in the coupling to potential and current are inherited from the signature of the Minkowski metric.
The coupling of the photon field to the external current $\jj$ is unitarily equivalent to including a classical vector potential in the minimal-coupling prescription~\cite{ruggenthaler2017-QEDFT}, but the form used here is preferable for a functional-theoretic formulation since the coupling then appears linearly.
Remember that we also needed to include a confining potential $\phiconf$ in order to break the symmetry of free space and guarantee the existence of a ground state for every choice of $\phi\in\potSpace$.
We do not include a fixed current in the same way (although this would be technically possible and allows to include, e.g., spatially constant magnetic fields or to approximate an external heat source).
The existence of a bound state in the usual Schrödinger equation allows us to show the existence of a ground-state also for the Pauli--Fierz Hamiltonian without an additional infrared cutoff~\cite{Griesemer2001}.
In addition, the fixed confining potential allows us to include singular Coulomb potentials from clamped nuclei that are themselves not elements of the space $\potSpace$. Further, we notice that the formulation is invariant under shifts of the confining potential by any element from $\potSpace$, because any such shift can be absorbed into the potential $\phi$.

We denote the Hamiltonian without additional sources, so $\phi=0$ and $\jj=0$ but still including the confining potential $\phiconf$, as $\hat{H}_{0}^{\Lambda}$.
The expectation value of $\hat{H}^{\Lambda}_{\phi,\jj}$ with respect to any ensemble state $\Gamma$ can then be evaluated as
\begin{equation}\label{eq:trace-H}\begin{aligned}
    \trace (\hat{H}^{\Lambda}_{\phi,\jj} \Gamma ) = \trace (\hat{H}^{\Lambda}_{0} \Gamma ) &+ \int \trace(\hat\rho(\rr)\Gamma) \phi(\rr) \,\rmd\rr \\
    &- \int \trace(\hAA^\Lambda(\rr)\Gamma) \jj(\rr) \,\rmd\rr,
\end{aligned}\end{equation}
where we separated off the universal part of the Hamiltonian from the parts that vary in $(\phi,\jj)\in\X^*$.

The dependence on the cutoff $\Lambda$ exactly corresponds to the adiabatic connection via interaction strength~\cite{Savin2003-AC} with the same zero-interaction (decoupling) limit $\Lambda\to 0$ that we will later use as the auxiliary Kohn--Sham system (Sec.~\ref{sec:renorm-ks}).
This holds since for $\Lambda=0$ there is no Coulomb interaction between the particles, $w^{\Lambda=0}_\mathrm{C}=0$. Furthermore, the quantized field then neither couples to the particles nor to the external current $\jj$, because the form factor vanishes and thus $\hAA^{\Lambda=0} = 0$.
In this limit, the non-interacting non-coupled system without external sources described by the Hamiltonian $\hat{H}_{0}^{\Lambda=0}$ will take the role of a reference system.
The solution of this system thus reduces to a one-particle Schr\"{o}dinger equation, as it is known from the Kohn--Sham method.
Instead of the bare mass $m^{\Lambda>0}$ we must use the \emph{observable} mass $m^{\Lambda=0}$ that corresponds to observations on single particles without electromagnetic self-interaction. Besides the fact that, like this, a Kohn--Sham system with a \emph{single electron} will have the correct observable free-space dispersion relation, this choice also guarantees that the \emph{non-interacting many-particle} Kohn--Sham system exhibits the correct plasma oscillations according to Eq.~\eqref{eq:Mie_frequency}. We discuss the importance of such many-body renormalization conditions in Sec.~\ref{sec:renorm-ks}.
Then, solving the reference Hamiltonian $\hat{H}_{0}^{\Lambda=0}$ with just the confining potential $\phiconf$, we arrive at an uncorrelated $N$-particle wave function and the vacuum state for the photon part in the ground state. We use this ground state to define reference quantities,
\begin{align}
    \label{eq:rho-ref-def}
    \rhoref(\rr) &= \trace(\hat\rho(\rr)\Gamma_0) \leq 0,\\
    \label{eq:A-ref-def}
    \AAref(\rr) &= \trace(\hAA^\Lambda(\rr)\Gamma_0) = 0 \quad\text{for any}\;\Lambda.
\end{align}
This $\rhoref$ fixes the charge sector (of total charge $-Ne$) and can be used as a common reference point for all particle densities in the following.
Since it is the non-interacting ground-state solution to just the confining potential $\phiconf$ with no additional potential $\phi$ added, it clearly depends on the choice of $\phiconf$, but as noted before, this potential can always be modified by any element of $\potSpace$ and consequently also $\rhoref$ can be adjusted this way.
Considering the physics relative to a reference system will be crucial when considering renormalization later in Sec.~\ref{sec:renorm-ks} and it is further a standard strategy in quantum field theory and related fields of research~\cite{thirring2002quantum, Ruetsche_2003}.

Coming back to the expectation value of the Hamiltonian with respect to an arbitrary state $\Gamma$ with finite energy in Eq.~\eqref{eq:trace-H}, we can similarly define reduced quantities that already describe how the system couples to $(\phi,\jj)\in\X^*$,
\begin{align}
    \label{eq:rho-exp-val}
    \rho(\rr) &= \trace(\hat\rho(\rr)\Gamma)\leq 0, \\
    \label{eq:A-exp-val}
    \AA(\rr) &= \trace(\hAA^\Lambda(\rr)\Gamma).
\end{align}
The SI units for these quantities are the same as for the charge density and vector potential after Eq.~\eqref{eq:Poisson-a}.
We now want to verify that these quantities always fit the space $\X$ when they correspond to a finite-energy state $\Gamma$, and that they are thus exactly from the dual space of $\X^*\ni(\phi,\jj)$. This dual structure will be later decisive for the formulation as a functional theory. For $\AA$ we have to check if all components of the vector potential have finite norm as given in Eq.~\eqref{eq:pot-norm}, i.e., we require
\begin{equation}
    \sum_{k=1}^3 \int |\nabla A_k(\rr)|^2 \rmd\rr < \infty.
\end{equation}
But with partial integration and again using the vector-Laplacian identity together with Coulomb gauge ($\nabla\cdot\AA=0$) this condition is equivalent to
\begin{equation}\label{eq:nablaXAestimate}
    \int |\nabla\times\AA(\rr)|^2 \rmd\rr < \infty,
\end{equation}
which equals finite magnetic field energy. But now we can apply the basic operator relation for expectation values $\langle\hat O\rangle^2 \leq \langle\hat O^2\rangle$ that equally applies for ensemble states to Eq.~\eqref{eq:nablaXAestimate} and Eq.~(13.75) from \citet{spohn-book} to get an estimate in terms of $\langle\hat H_\mathrm{field}\rangle$, which must be finite for any finite-energy state.
This shows $\AA\in\potSpaceVec$ for an expectation value like in Eq.~\eqref{eq:A-exp-val}.

To show also that $\rho\in\densSpace$ with a density like in Eq.~\eqref{eq:rho-exp-val}, we note that for any state with finite (kinetic) energy we have $\rho \in L^1\cap L^3$, which gives the standard setting for DFT due to \citet{Lieb1983}.
We use the continuous embedding $L^1\cap L^3 \subseteq L^{6/5}$ \cite[remark after Th.~3.9]{Lieb1983} which transforms to $L^{3/2}+L^\infty \supseteq L^{6}$ on the dual side. Together with the result $\potSpace\subseteq L^6$ from Eq.~\eqref{eq:hom-Sobolev-embedding}, which yields $\densSpace\supseteq L^{6/5}$ by duality, this gives the chain of continuous embeddings
\begin{equation}\label{eq:embedding}
    L^1\cap L^3 \subseteq L^{6/5} \subseteq \densSpace.
\end{equation}
We thus also have $\rho\in\densSpace$ as required. On the dual side Eq.~\eqref{eq:embedding} translates to
\begin{equation}\label{eq:embedding-dual}
    L^{3/2}+L^\infty \supseteq L^{6} \supseteq \potSpace.
\end{equation}
So taking a potential $\phi\in\potSpace$ means that it is automatically in $L^{3/2}+L^\infty$. The KLMN theorem~\cite{reed-simon-2,spohn-book} then guarantees a self-adjoint and bounded-below Hamiltonian with stable \emph{form} domain when adding any potential $\phi\in\potSpace$ to the fixed $\phiconf$. A formulation based on quadratic forms is sufficient here, since in functional theories we are only concerned with expectation values.
The same theorem allows to show that the coupling to the external current $\jj\in\densSpaceVec$ can be added as well. For this let $\jj=\aa^*=-\mu_0^{-1}\Delta_\mathrm{vec}\aa=\mu_0^{-1}\nabla\times\nabla\times\aa$ with $\aa\in\potSpaceVec$ and use partial integration to show
\begin{equation}
    \int \hAA^\Lambda(\rr) \cdot \jj(\rr) \rmd \rr = \frac{1}{\mu_0}\int (\nabla\times\hAA^\Lambda(\rr)) \cdot (\nabla\times\aa(\rr)) \rmd \rr.
\end{equation}
For any $a\in\R$ the simple inequality $(ax-y/a)^2=a^2x^2+y^2/a^2-2xy\geq 0$ then allows the estimate
\begin{equation}
    \left| \int \! \hAA^\Lambda(\rr) \cdot \jj(\rr) \rmd \rr \right| \leq \frac{a^2}{2\mu_0}\int\! |\nabla\times\hAA^\Lambda(\rr)|^2\rmd\rr + \frac{\|\aa\|_{\potSpaceVec}^2}{2a^2}.
\end{equation}
This means the condition for the KLMN theorem~\cite[Th.~X.17]{reed-simon-2} are fulfilled if we choose $a<1$ and estimate through $\hat H_\mathrm{field}$.

We can now put the energy-expectation value of Eq.~\eqref{eq:trace-H} in a short form using a dual pairing between $(\phi,\jj)\in\X^*$ and $(\rho,\AA)\in\X$ (with an extra minus sign in the current coupling),
\begin{equation}\label{eq:trace-H-dual}\begin{aligned}
    \trace (\hat{H}^{\Lambda}_{\phi,\jj} \Gamma ) &= \trace (\hat{H}^{\Lambda}_{0} \Gamma ) + \langle \phi,\rho \rangle - \langle \jj,\AA\rangle.
\end{aligned}\end{equation}
Note that this changes the order in how we write the dual pairing between potentials and densities when compared to Eq.~\eqref{eq:P-dual-pairing} and that we prefer this order from here on.
In summary, for every choice of $(\Lambda,m^{\Lambda})$ as well as $(\phi,\jj)\in\X^*$ the Pauli--Fierz Hamiltonian has a ground state $\Gamma$ that yields expectation values $(\rho,\AA)$ as in Eqs.~\eqref{eq:rho-exp-val} and \eqref{eq:A-exp-val}. We will now reformulate this ground-state setting  into a non-linear theory in terms of the reduced quantities $(\rho,\AA)$ that entirely replace the wave function in a functional-theoretic formulation.     

\section{Functional-theoretic reformulation}
\label{sec:df-formulation}

We fix $(\Lambda,m^{\Lambda})$ for our theory, as well as the confining potential $\phiconf$, but let $(\phi,\jj)\in\X^*$ vary as an element of the space introduced in Sec.~\ref{sec:Maxwellhilbert}. Then, by solving the Pauli--Fierz Hamiltonian for an (ensemble) ground state $\Gamma$, we have a mapping from the sources (external quantities) $(\phi,\jj)$ to the expectation values (internal quantities) $(\rho,\AA)$ in the ground state as in Eqs.~\eqref{eq:rho-exp-val} and \eqref{eq:A-exp-val},
\begin{align}
(\phi,\jj) \mapsto \Gamma \mapsto ( \rho,\AA ).
\end{align}
We always have $(\rho,\AA)\in\X$ by the results of Sec.~\ref{sec:qed-and-reference}.
Just like in the standard functional-theoretic reformulation of the Pauli--Fierz ground-state problem as \ac{QEDFT}~\cite{ruggenthaler2017-QEDFT,ruggenthaler2023understanding,penz2023structure}, we can now set up a constrained-search functional that expresses the \emph{internal energy} of the system,
\begin{equation}\label{eq:QEDFTfunctional}
F^{\Lambda}(\rho,\AA) = \inf_{\Gamma' \mapsto (\rho,\AA)} \trace (\hat{H}_0^{\Lambda} \Gamma' ).
\end{equation}
Here, the shorthand notation $\Gamma'\mapsto(\rho,\AA)$ is used to express that we consider only states $\Gamma'$ that map to $(\rho,\AA)$ by Eqs.~\eqref{eq:rho-exp-val} and \eqref{eq:A-exp-val}.
We will generally use primed quantities for variation to not confuse them with the optimizers of such a variational problem.
In the context of \ac{QEDFT}, this is called the \emph{universal functional} since it is independent of the specific choice of system parameters and just includes the fixed part of the Hamiltonian.
This functional is convex in both arguments by the same reasoning as in \citet[Eq.~(4.3)]{Lieb1983}, but critically also needs to be lower semicontinuous w.r.t.\ to the topology of the density space $\X$ for a functional-theoretic formulation. This lower semicontinuity together with the fact that $\X$ is a Hilbert space requires a setting where every choice of $(\phi,\jj)\in\X^*$ supports a ground state~\cite{MY-Perspective}, which is indeed fulfilled here through the introduction of the confining potential $\phiconf$. It must be stressed that at the present moment a mathematical proof for the lower semicontinuity of $F^{\Lambda}(\rho,\AA)$ is still missing, but was achieved recently in a similar functional-analytic setting on a periodic domain~\cite{MY-periodic}.

The ground-state energy according to the Rayleigh--Ritz principle, a concave functional in $(\phi,\jj)$~\cite[like in Th.~3.1(i)]{Lieb1983}, then immediately becomes the Legendre--Fenchel transformation of $F^{\Lambda}$ when using the expression from Eq.~\eqref{eq:trace-H-dual} for the expectation value of the Hamiltonian,
\begin{equation}\label{eq:E-LF}
\begin{aligned}
E^{\Lambda}(\phi,\jj) &= \inf_{\Gamma'} \trace (\hat{H}^{\Lambda}_{\phi,\jj} \Gamma' ) \\
&= \inf_{(\rho',\AA')}\left\{ F^{\Lambda}(\rho',\AA') + \langle \phi,\rho' \rangle - \langle \jj,\AA' \rangle \right\}.
\end{aligned}
\end{equation}
A minimizer $(\rho,\AA)\in\X$ of this variational problem satisfies
\begin{equation}\label{eq:F-subdiff}
    (-\phi,\jj)\in\subdiff F^{\Lambda}(\rho,\AA).
\end{equation}
Instead of a functional derivative ``$\rmd$'' we must employ the subdifferential ``$\subdiff$'' here, that is the set of all slopes of tangents at $(\rho,\AA)$ lying below the convex functional. Instead of a single derivative, we then get a whole set of (sub)gradients in $\X^*$.
Note that although hidden in the notation, these definitions differ slightly from previous formulations of \ac{QEDFT} in that the Hamiltonian always includes the confining potential $\phiconf$ and any chosen $\phi$ is added to it. At the present stage, these are just redefinitions of known structures.
These structures are already very powerful and allow to construct \ac{QEDFT} to perform ab-initio \ac{QED} simulations~\cite{ruggenthaler2023understanding,bauman2025perspective}. Nevertheless, there are some known mathematical subtleties that are hard to overcome in a formulation only based on the electron-photon quantities. On the one hand, the previous formulation of \ac{QEDFT} inherits differentiability issues regarding $F^{\Lambda}(\rho,\AA)$ due to inconvenient topological properties of the underlying function spaces from standard \ac{DFT}~\cite{Lammert2007,Penz2023-PartI,MY-Perspective}, and on the other hand, suffers from the renormalization problems of \ac{QED}~\cite{ruggenthaler2023understanding}.

However, if we want to describe and capture the physics of the \emph{total} system and not of the (only theoretically distinct) electron-photon subsystem alone, we should also consider a functional reformulation of the total system including $(\phi,\jj)$ not just as parameters of the Hamiltonian but as \emph{physical} quantities. This is the central idea behind the ensuing regularizing transformation. The duality properties of the spaces devised for solutions to the Maxwell equations in Sec.~\ref{sec:Maxwellhilbert} allow us to transform $(\phi,\jj)\in\X^*$ canonically into a charge density and vector potential pair $(\phi^*,\jj^*)\in\X$. Remember that this means that $(\phi,\jj^*)$ solves the Maxwell equations for external charge distribution and current $(\phi^*,\jj)$. For a ground state $\Gamma$ of $\hat H^\Lambda_{\phi,\jj}$ with $\Gamma\mapsto(\rho,\AA)$, we can then consider the \emph{mixed} charge density and vector potential
\begin{align}\label{eq:def-mixed}
\bar\rho = \rho - \phi^{*},\quad \bAA = \AA + \jj^{*}.
\end{align}
Since for every $(\phi,\jj)\in\X^*$ it holds $(\phi^*,\jj^*)\in\X$ by Maxwell-duality, we importantly have that
\begin{equation}
    (\bar\rho,\bAA)\in\X,
\end{equation}
so the newly defined quantities are in the same space as $(\rho,\AA)\in\X$.
The new mixed density is thus defined as the \emph{difference} between the charge density of the quantum system and the external charge. Hence, this does \emph{not} give the total density of the system, internal plus external, and the reason for considering the difference instead is the following. A given $\phi$ will attract negative charges to regions where the external charge density given by $\phi^*$ is positive and repel them from where it is negative. Thus, $\rho+\phi^*$ dilutes the effect that the $\phi$ has on $\rho$, while in $\bar\rho$ this effect gets amplified. This is important since in a functional theory we want to be able to uniquely map from the new variables to the potentials. Indeed, this will follow as a mathematical fact when changing to the new variables and defining an adapted universal functional below. The same is true for the new vector potential that sums the expectation value of the quantum system with the classical vector potential that corresponds to the current $\jj$ via the Maxwell equations. It comes with a different sign due to the minus in the coupling between $\hAA^\Lambda$ and $\jj$, while it is a positive sign between $\hat\rho$ (which already includes the negative charge) and $\phi$ in the Hamiltonian of Eq.~\eqref{eq:H}. The simpler reason behind those different signs is that like charges repel each other, yet parallel currents attract themselves. Consequently, the different signs in Eq.~\eqref{eq:def-mixed} are not a matter of choice but a physical fact. While we treat the new variables $(\bar\rho, \bAA)$ as a purely theoretical tool at the moment, they will receive a physical interpretation in Sec.~\ref{sec:interpret}.

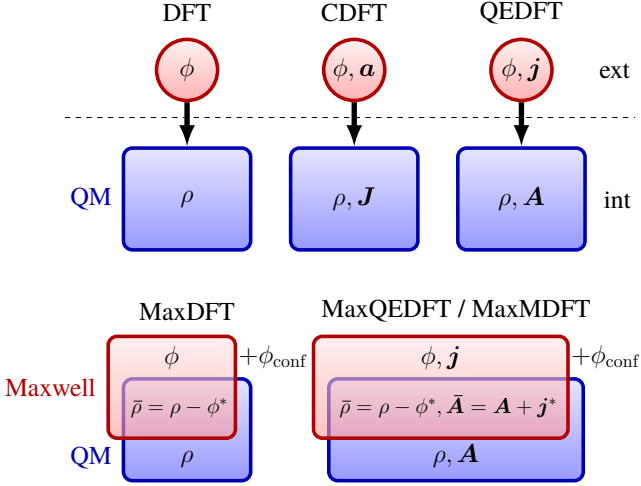
\begin{figure}
\resizebox{1\columnwidth}{!}{%
\begin{tikzpicture}[
  node font=\Large,
  box/.style={
    rectangle, rounded corners=5pt,
    draw=blue!70!black, line width=2pt,
    top color=blue!10, bottom color=blue!40,
    minimum width=2.5cm, minimum height=2cm,
    inner sep=5pt, align=center
  },
  boxalt/.style={
    rectangle, rounded corners=5pt,
    draw=red!70!black, line width=2pt,
    top color=red!10, bottom color=red!40,
    minimum width=2.5cm, minimum height=2cm,
    inner sep=5pt, align=center,
    fill opacity=0.6
  },
  circ/.style={
    circle, draw=red!70!black, line width=2pt,
    top color=red!5, bottom color=red!30,
    minimum size=1.2cm,
    inner sep=0pt, align=center
  },
  arrow/.style={
    -latex, line width=3pt
  }
]

\draw[dashed] (-2.4,1.6) -- (8.8,1.6);
\node[box] (box1) {$\rho$};
\node[circ] (circ1) at (box1.north) [yshift=1.5cm] {$\phi$};
\draw[arrow] (circ1.south) -- (box1.north);
\node at (circ1.north) [yshift=.5cm] {DFT};
\node[box] (box2) at (box1.east) [xshift=2cm] {$\rho,\JJ$};
\node[circ] (circ2) at (box2.north) [yshift=1.5cm] {$\phi,\aa$};
\draw[arrow] (circ2.south) -- (box2.north);
\node at (circ2.north) [yshift=.5cm] {CDFT};
\node[box] (box3) at (box2.east) [xshift=2cm] {$\rho,\AA$};
\node[circ] (circ3) at (box3.north) [yshift=1.5cm] {$\phi,\jj$};
\draw[arrow] (circ3.south) -- (box3.north);
\node at (circ3.north) [yshift=.5cm] {QEDFT};
\node at (circ3.east) [xshift=1.2cm,align=center] {ext};
\node at (box3.east) [xshift=.6cm] {int};
\node[text=blue!70!black] at (box1.west) [xshift=-.6cm] {QM};

\node[box] (box41) at (box1.south) [yshift=-3.5cm] {};
\node[boxalt] (box42) at (box41.north) [yshift=-.2cm,xshift=-.3cm] {};
\node[above=-1.95cm of box41] {$\rho$};
\node[above=-0.85cm of box42] {$\phi$};
\node[above=-1.8cm of box42,xshift=.15cm] {\large $\bar\rho=\rho-\phi^*$};
\node at (box41.north) [yshift=1.3cm] {MaxDFT};
\node[text=blue!70!black] at (box41.west) [xshift=-.6cm,yshift=-.55cm] {QM};
\node[text=red!70!black] at (box42.west) [xshift=-1.1cm] {Maxwell};

\node[box,minimum width=5cm] (box51) at (box2.south) [yshift=-3.5cm,xshift=2cm] {};
\node[boxalt,minimum width=5cm] (box52) at (box51.north) [yshift=-.2cm,xshift=-.3cm] {};
\node[above=-1.95cm of box51] {$\rho,\AA$};
\node[above=-0.85cm of box52] {$\phi,\jj$};
\node[above=-1.8cm of box52,xshift=.15cm] {\large $\bar\rho=\rho-\phi^*, \bAA=\AA+\jj^*$};
\node at (box51.north) [yshift=1.3cm] {MaxQEDFT / MaxMDFT};
\node at (box42.east) [xshift=.71cm,yshift=.6cm,align=center] {$+\phiconf$};
\node at (box52.east) [xshift=.71cm,yshift=.6cm,align=center] {$+\phiconf$};

\end{tikzpicture}
}
\caption{Comparison of standard functional theories, where external source quantities serve as a parametrization of the ground-state, with their Maxwell-regularized counterparts, where the sources receive a physical description and are combined with the internal quantities to form the mixed ones. Note that current \ac{DFT} ({\smaller CDFT}) has no direct counterpart, but {\smaller MaxMDFT} is instead derived in Sec.~\ref{sec:MaxMDFT} as an alternative theory for including magnetic fields.}
\label{fig:mappings}
\end{figure}

Giving the sources $(\phi,\jj)$ within the Hamiltonian a separate physical reality, we should also consider their electromagnetic energy. We thus need to add this energy as an \emph{externalized cost}, given by Eq.~\eqref{eq:def-F-elm}, to the universal functional that represents the internal energy of the system.
This defines the new functional,
\begin{equation}\label{eq:F-regularized}
    \bar{F}^{\Lambda}(\bar\rho,\bAA) = F^{\Lambda}(\rho, \AA) + \Eelm(\phi,\jj).
\end{equation}
An equivalent definition uses variation over an auxiliary pair $(\rho',\AA')$ to write
\begin{equation}\label{eq:F-regularized-alt}\begin{aligned}
    \bar{F}^{\Lambda}(\bar\rho,\bAA) = \!\!\inf_{(\rho', \AA')}\left\{F^{\Lambda}(\rho', \AA') + \frac{1}{2} \| (\bar\rho-\rho',\bAA-\AA') \|_{\X}^2\right\}. 
\end{aligned}\end{equation}
Like in Eq.~\eqref{eq:E-LF} any minimizer $(\rho,\AA)$ must fulfill $(\bar\rho-\rho,\bAA-\AA)^*\in\subdiff F^{\Lambda}(\rho, \AA)$, since the derivative of the square norm in $\X$ just yields the Maxwell-duality mapping as in Eq.~\eqref{eq:square-deriv-X}.
By Eq.~\eqref{eq:F-subdiff} this gives us $(\bar\rho-\rho)^*=-\phi$ and $(\bAA-\AA)^*=\jj$ as a solution. This shows that the variational formulation just retrieves Eq.~\eqref{eq:F-regularized} again, where the arguments are connected as in Eq.~\eqref{eq:def-mixed}.
Importantly, as Eq.~\eqref{eq:F-regularized-alt} shows, the transformation from $F^\Lambda$ to $\bar F^\Lambda$ exactly corresponds to the \emph{Moreau--Yosida regularization} (or Moreau envelope)~\cite{moreau1965,zalinescu2002,Bauschke-Combettes,MY-Perspective} and it has numerous extremely useful properties~\cite[Th.~18.15]{Bauschke-Combettes}.
It usually carries a single index $\varepsilon$ in the notation that appears here in the form of $\epsilon_0$ and $\mu_0^{-1}$, but we have absorbed them into the definition of the norm of $\X$ and thus they do not show up explicitly.
In the context of Moreau--Yosida regularization, the minimizer $(\rho,\AA)$ in Eq.~\eqref{eq:F-regularized-alt} is called the proximal point~\cite[Def.~12.23]{Bauschke-Combettes} of the mixed quantities $(\bar\rho,\bAA)$, a concept that has its very own domain of applications~\cite{Parikh2014}.
Clearly the most important consequence is that $\bar F^\Lambda$ is now \emph{Fr\'{e}chet differentiable} with a Lipschitz-continuous derivative (with constant 1).
In full analogy to Eq.~\eqref{eq:E-LF} we define the Legendre--Fenchel transformation of $\bar F^\Lambda$ as $\bar E^\Lambda$. We can then put in the definition for $\bar F^\Lambda$ from Eq.~\eqref{eq:F-regularized-alt}, swap the two infima, and get
\begin{equation}\label{eq:regularizedE}\begin{aligned}
    &\bar E^\Lambda(\phi,\jj) = \inf_{(\bar\rho',\bAA')}\left\{ \bar F^\Lambda(\bar\rho',\bAA') + \langle \phi,\bar\rho' \rangle - \langle \jj,\bAA'\rangle \right\} \\
    &= \inf_{(\rho', \AA')}\bigg\{F^{\Lambda}(\rho', \AA') + \inf_{(\bar\rho',\bAA')}\bigg\{\frac{1}{2} \| (\bar\rho'-\rho',\bAA'-\AA') \|_{\X}^2 \\
    &\quad + \langle \phi,\bar\rho' \rangle - \langle \jj,\bAA'\rangle \bigg\}\bigg\} \\
    &= \inf_{(\rho', \AA')}\left\{F^{\Lambda}(\rho', \AA') + \langle\phi,\rho'\rangle - \langle\jj,\AA'\rangle \right\} - \frac{1}{2}\|(\phi,\jj)\|^2_{\X^*}\\
    &= E^\Lambda(\phi,\jj) - \frac{1}{2}\|(\phi,\jj)\|^2_{\X^*} = E^\Lambda(\phi,\jj) - \Eelm(\phi,\jj).
\end{aligned}\end{equation}
Here, the inner infimum in the second line was evaluated with differentiation, where by Eq.~\eqref{eq:square-deriv-X} the derivative of the square norm in $\X$ just yields the Maxwell-duality mapping again, $(\bar\rho-\rho')^*=-\phi$ and $(\bAA-\AA')^*=\jj$. This explains the choice of variables in Eq.~\eqref{eq:def-mixed} and reveals the simple relation between $\bar E^\Lambda$ and $E^\Lambda$ that is exactly through subtraction of the electromagnetic field energy.
By Legendre--Fenchel back-transformation, we recover
\begin{equation}\label{eq:Ebar-LF}
    \bar F^\Lambda(\bar\rho,\bAA) = \sup_{(\phi',\jj')} \left\{ \bar E^\Lambda(\phi',\jj') - \langle \phi',\bar\rho \rangle + \langle \jj',\bAA\rangle \right\}.
\end{equation}
We can thus conclude that by considering the total system in terms of mixed quantities and including the field energy of $(\phi,\jj)$, we automatically switch to a regularized theory with a differentiable universal functional $\bar F^\Lambda$ that in the theory unfolding here receives the attribute \emph{Maxwell-regularized}. Fig.~\ref{fig:mappings} illustrates a comparison between the standard functional theories and their Maxwell-regularized counterparts.

The differentiability of the new functional means that the variational problem in the first line of Eq.~\eqref{eq:regularizedE} can be uniquely solved by
\begin{equation}\label{eq:nabla-F}
    (-\phi,\jj) = \rmd\bar F^\Lambda(\bar\rho,\bAA).
\end{equation}
Here, the Fréchet derivative of $\bar F^\Lambda$ evaluated at a point $(\bar\rho,\bAA)$ always returns a linear functional on the space $\X$ and thus an element in $\X^*$, $(-\phi,\jj)$ in this case.
We prefer this notation for the functional derivative over writing $\delta \bar F^\Lambda/\delta \bar\rho$ and $\delta \bar F^\Lambda/\delta\bAA$ which is usually found in a physics context, since it automatically encompasses all components, clearly describes an element in $\X^*$ when evaluated at a point in $\X$, and further avoids ambiguities that can arise in the other notation.
Equation~\eqref{eq:nabla-F} now gives the parameters $(\phi,\jj)$ for the Hamiltonian $\hat H^\Lambda_{\phi,\jj}$ that exactly yields $(\bar\rho,\bAA)\in\X$ from the ground-state solution. The usual $v$-representability problem~\cite{dreizler-gross-book,Englisch_1983,Garrigue2022,Trushin2024} does not show up in this case, since representability is now for $(\bar\rho,\bAA)=(\rho-\phi^*,\AA+\jj^*)$ instead of only the expectation-value quantities $(\rho,\AA)$ of the quantum system (this feature was already noted before~\cite[Sec.~IX]{Penz2023-PartI}).
It further means that we automatically have the result of the Hohenberg--Kohn theorem, since every $(\bar\rho,\bAA)$ yields a \emph{unique} external $(\phi,\jj)\in\X^*$.
In particular, for the mixed density we have $\bar\rho = \rho-\phi^*$, which with the addition of $-\phi^*=\epsilon_0\Delta\phi$ is not limited to purely negative charge densities any more. This already alleviates certain representability issues. 
But more significantly, the $\bar\rho$ always also includes the external potential $\phi$. If a slight variation $\delta\rho$ requires a large change of potential $\delta\phi$, then this will automatically correspond to a large variation in $\delta\bar\rho$ as well. Further, the variations of $\phi$ and $\rho$ will not cancel each other in $\delta\bar\rho=\delta\rho-\delta\phi^*$ as detailed after Eq.~\eqref{eq:def-mixed}. These features all relate to the differentiability of $\bar F^\Lambda$ and the Lipschitz-continuity of $\rmd\bar F^\Lambda$.

\begin{figure}
\resizebox{.9\columnwidth}{!}{%
\begin{tikzpicture}[
  node font=\Large,
  box/.style={
    rectangle, rounded corners=5pt,
    draw=orange!70!black, line width=2pt,
    top color=orange!10, bottom color=orange!40,
    minimum width=1.5cm, minimum height=1.5cm,
    inner sep=5pt, align=center
  },
  smbox/.style={
    rectangle, rounded corners=5pt,
    draw=black, line width=2pt,
    minimum width=.75cm, minimum height=.75cm,
    inner sep=5pt, align=center
  },
  circ/.style={
    circle, draw=red!70!black, line width=2pt,
    top color=red!5, bottom color=red!30,
    minimum size=1.7cm,
    inner sep=0pt, align=center
  },
  circalt/.style={
    circle, draw=blue!70!black, line width=2pt,
    top color=blue!5, bottom color=blue!30,
    minimum size=1.7cm,
    inner sep=0pt, align=center
  },
  circviol/.style={
    circle, draw=violet!70!black, line width=2pt,
    top color=violet!10, bottom color=violet!40,
    minimum size=1.7cm,
    inner sep=0pt, align=center
  },
  arrow/.style={
    -latex, line width=3pt
  }
]

\node[circ] (circ11) {$\phi,\jj$};
\node[box] (box11) at (circ11.east) [xshift=2.1cm] {$\supdiff E^\Lambda$};
\draw[arrow] (circ11.east) -- (box11.west);
\node[circalt] (circ12) at (box11.east) [xshift=2.1cm] {$\rho,\AA$};
\draw[arrow] (box11.east) -- (circ12.west);

\node[smbox] (minusbox) at (circ12.east) [xshift=1.2cm] {$\bm{\mp}$};
\draw[arrow] (circ12.east) -- (minusbox.west);

\node[circviol] (circ13) at (minusbox.east) [xshift=1.7cm] {$\bar\rho,\bAA$};
\draw[arrow] (minusbox.east) -- (circ13.west);
\node[box] (box12) at (box11.south) [yshift=-1cm] {$()^*$};
\node[align=center] at (box12.south) [yshift=-.8cm] {Maxwell\\[-.2em]duality};
\draw[arrow,rounded corners=10pt] (circ11.south) |- (box12.west);
\draw[arrow,rounded corners=10pt] (box12.east) -| (minusbox.south);

\node[box] (box01) at (box11.north) [yshift=1cm] {$\rmd \bar F^\Lambda$};
\draw[arrow,rounded corners=10pt] (circ13.north) |- (box01.east);
\draw[arrow,rounded corners=10pt] (box01.west) -| (circ11.north);
\end{tikzpicture}
}
\caption{The structure of \ac{MaxQEDFT} as an invertible input--output scheme with the formulation in terms of $(\phi,\jj)$ that enters the Pauli--Fierz Hamiltonian and the $(\rho,\AA)$ from expectation values of a ground-state solution. The result are the mixed quantities $(\bar\rho,\bAA)$ from Eq.~\eqref{eq:def-mixed} as a combination of internal and external quantities.}
\label{fig:io}
\end{figure}
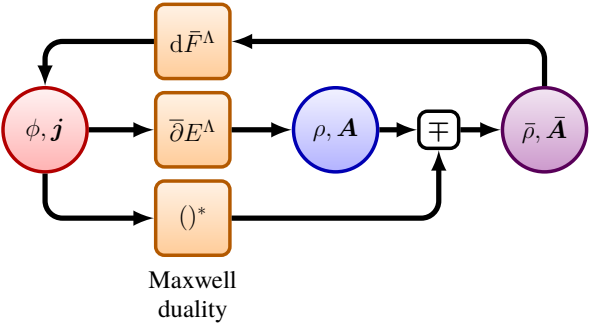

Solving instead the variational problem in Eq.~\eqref{eq:Ebar-LF} is equivalent to finding the inverse of Eq.~\eqref{eq:nabla-F} by the rules of convex analysis~\cite[Cor.~16.24]{Bauschke-Combettes}. The solution is given by the superdifferential of $\bar E^\Lambda$ at $(\phi,\jj)$ (defined in analogy to the subdifferential for a concave functional) that does not need to yield a unique $(\rho,\AA)$ (or $(\bar\rho,\bAA)$) and we write
\begin{equation}\label{eq:Aint-and-A-supdiff}
    (\rho,-\AA) \in \supdiff E^\Lambda(\phi,\jj) \; {\rm and} \; (\bar\rho,-\bAA)\in \supdiff \bar E^\Lambda(\phi,\jj).
\end{equation}
Since such a $(\rho,\AA)$ exactly yields the minimizer in Eq.~\eqref{eq:E-LF} for a given $(\phi,\jj)\in\X^*$, this corresponds to the ground-state solution of the Pauli--Fierz Hamiltonian. The possibility of degeneracy in the ground state then also explains the non-uniqueness, that expresses itself in $\supdiff E^\Lambda(\phi,\jj)$ and $\supdiff \bar E^\Lambda(\phi,\jj)$ being set-valued.
Yet, since the confining potential $\phiconf$ always guarantees existence of a ground state, those sets are never empty. The superdifferential of Eq.~\eqref{eq:regularizedE} yields right away
\begin{equation}\label{eq:diff-Ebar-E}
    \supdiff \bar E^\Lambda(\phi,\jj) = \supdiff E^\Lambda(\phi,\jj) - (\phi^*,\jj^*),
\end{equation}
which with Eq.~\eqref{eq:Aint-and-A-supdiff} leads back to the basic relation $(\bar\rho,\bAA)=(\rho-\phi^*,\AA+\jj^*)$.

This switch to mixed quantities now manifests itself in the basic mappings. In \ac{QEDFT} we map $(\phi,\jj) \mapsto (\rho,\AA)$, which means \emph{external} to \emph{internal} (see Fig.~\ref{fig:mappings}). In \ac{MaxQEDFT} we instead map $(\phi,\jj) \mapsto (\bar\rho,\bAA)$, which means \emph{external} to \emph{mixed}. Importantly, we are now also able to uniquely perform the inverse mapping $(\bar\rho,\bAA) \mapsto (\phi,\jj)$ via Eq.~\eqref{eq:nabla-F}. With \ac{MaxQEDFT} we thus also have realized a perfect \emph{control theory}, where any choice of mixed quantities can be achieved with a corresponding choice of potential and current. The whole structure is displayed in Fig.~\ref{fig:io}. We note that the whole quantum layer is fully compressed within the functionals defined in this section that encode the full information about the electron-photon system under influence of $(\phi,\jj)$ in the equilibrium. This includes all theoretical choices (quantum subsystem description, gauges, form of cutoff, etc.) that are hidden within the non-linear but well-behaved functionals, and we will use exactly this fact in Sec.~\ref{sec:renorm-ks} to consider the $\Lambda \rightarrow \infty$ limit. 
In this framework, the wave function (or density matrix) is reduced to a purely computational tool that enters only in the evaluation of the constrained-search functional to test out all possible configurations in the optimization. One can replace the unfeasible correlated electron-photon wave function by an auxiliary Kohn--Sham system as detailed in Sec.~\ref{sec:renorm-ks}. The exact same transformation to a Maxwell-regularized functional can also be performed in standard DFT by considering the energy of the electrostatic field $\frac{1}{2}\|\phi\|^2_\potSpace$. This setting is separately considered in Sec.~\ref{sec:MaxDFT}.
Yet, the nature of the new, mixed quantities $(\bar\rho,\bAA)$ still remains enigmatic at this point, but will be explored in the following.

\section{Emergent physical picture}
\label{sec:interpret}

At this stage we want to take a step back and try to understand the emerging physical picture. We have seen that the main point of the new approach is to take the energy of the external sources into account. Due to Maxwell duality, this is equivalent to the energy of the electromagnetical field that constitutes the environment. 
Thus besides guaranteeing finite energy for the quantum system, both, the system and the environment are being considered with finite energy content.
The consistent description of the system \emph{and} sources by quantities from the spaces $\X$ and $\X^*$, developed for the formulation of the Maxwell equations, will now allow us to capture the back-action of the quantum system on the source fields and answer the question on the interpretation of the mixed quantities.

The central object of \ac{MaxQEDFT} is the Maxwell-regularized universal functional $\bar{F}^{\Lambda}(\bar\rho,\bAA)$. At the solution point it is defined by Eq.~\eqref{eq:F-regularized}. It takes into account the internal energy of the quantum system describes by $(\rho,\AA)\in\X$ and the electromagnetic energy of the classical sources $(\phi,\jj)\in\X^*$ that generate this specific quantum state. The general form of the Maxwell-regularized universal functional, however, that is given in terms of the mixed quantities $(\bar \rho, \bAA)$ in Eq.~\eqref{eq:F-regularized-alt} allows us to find a physical interpretation for them. For this, we insert the constrained-search definition of the original functional from Eq.~\eqref{eq:QEDFTfunctional} and expand the square term to find
\begin{equation}\begin{aligned}
    \bar{F}^{\Lambda}(\bar\rho,\bAA) = \!\!\inf_{(\rho', \AA')}\bigg\{ & \inf_{\Gamma' \mapsto (\rho',\AA')} \trace (\hat{H}_0^{\Lambda} \Gamma' ) + \frac{1}{2} \| (\rho',\AA') \|_{\X}^2 \\
    &- \langle (\bar\rho,\bAA)^*, (\rho',\AA')\rangle \bigg\} + \frac{1}{2} \| (\bar\rho,\bAA) \|_{\X}^2. 
\end{aligned}\end{equation}
Now, the infima can be combined again if we vary over all density matrices $\Gamma'$ and instead of $(\rho',\AA')$ consider $(\rho_{\Gamma'},\AA_{\Gamma'})$ that denotes the expectation values of the density and vector potential with respect to the state $\Gamma'$. We arrive at
\begin{equation}\label{eq:Fbar-variational}
\begin{aligned}
    \bar{F}^{\Lambda}(\bar\rho,\bAA) = \inf_{\Gamma'}\bigg\{ & \trace (\hat{H}_0^{\Lambda} \Gamma' ) + \frac{1}{2} \| (\rho_{\Gamma'},\AA_{\Gamma'}) \|_{\X}^2 \\
    &- \langle (\bar\rho,\bAA)^*, (\rho_{\Gamma'},\AA_{\Gamma'})\rangle \bigg\} + \frac{1}{2} \| (\bar\rho,\bAA) \|_{\X}^2
\end{aligned}
\end{equation}
and variation over the quantum state then yields a non-linear Schrödinger equation that needs to be evaluated for its ground state $\Gamma$. Here, the first $\tfrac{1}{2}$ gets canceled from the differentiation of the square term and thus the Hamiltonian gets modified by an effective potential and current $(\rho_{\Gamma}-\bar\rho,-\AA_{\Gamma}+\bAA)^*$ (note the different signs in the coupling as in Eq.~\eqref{eq:E-LF}). Since this term already includes the ground-state solution $\Gamma$, the equation becomes non-linear and the solution must be found self-consistently, very similar to a Hartree scheme but still with Coulomb interactions present in $\hat H_0^\Lambda$ if $\Lambda>0$. If, on the other hand, $\Lambda=0$ then this exactly corresponds to a Hartree problem and we thus have $\bar{F}^{0}(\bar\rho,\bAA) = E^{\rm H}(-\bar\rho^*,\bAA^*) + \frac{1}{2} \| (\bar\rho,\bAA) \|_{\X}^2$. (Using the Hartree method to map $\bar\rho\mapsto\rho$ is equivalent to the proximal-point iteration~\cite[Sec.~IV.A]{MY-Perspective}.) In any case, at the solution point we then find $(\rho_{\Gamma}-\bar\rho,-\AA_{\Gamma}+\bAA)^* = (\phi,\jj)$ according to Eq.~\eqref{eq:F-regularized}, just as in our fundamental relation for the mixed quantities in Eq.~\eqref{eq:def-mixed}, so this ground state is the same as the one from the \emph{linear} Pauli--Fierz theory for given $(\phi,\jj)$. But instead of just using fixed parameters $(\phi,\jj)$, these external sources now depend on the ground state of the quantum system self-consistently. The \emph{\emph{total}} system features \emph{bare} external sources $(-\bar\rho,\bAA)^*$ that get modified by the potential and current produced by the quantum system via the Maxwell equations (represented by the Maxwell duality operation $()^*$), $(\rho_{\Gamma},-\AA_{\Gamma})^*$ and combine to yield a \emph{screened} potential and current $(\phi,\jj)$.

We will now use the language of the macroscopic Maxwell equations~\cite{jackson-book} to clarify this back-action of the quantum system on the bare sources $(-\bar \rho, \bAA)^*$. If we introduce the following vector fields,
\begin{alignat}{2}
    & \EE = -\nabla\phi, \quad && \BB = \nabla\times\jj^*, \label{eq:MacroMax-E-B} \\
    & \DD = \eps_0\nabla\bar\rho^*, \quad && \HH = \mu_0^{-1}\nabla \times \bAA, \label{eq:MacroMax-D-H} \\
    & \PP = \eps_0\nabla\rho^*, \quad && \MM = -\mu_0^{-1}\nabla \times \AA, \label{eq:MacroMax-P-M}
\end{alignat}
we can then easily verify the relations from Eq.~\eqref{eq:def-mixed} via the usual macroscopic Maxwell equations,
\begin{equation}
    \DD=\eps_0\EE+\PP,\quad \HH=\mu_0^{-1}\BB-\MM.
\end{equation}
From the definition of bound density and current we get $\rhob = -\nabla\cdot\PP = \rho$ and $\jjb=\nabla\times\MM=-\AA^*$ by the Maxwell duality. This reflects the polarization of the confined (bound) quantum system. In case of the vector potential $\AA$, this \emph{vacuum magnetization} is the expectation value of the field operator and consequently yields the total field that originates from coupling to the external source $\jj$ \emph{and} the quantum matter system. In contrast to the bound quantities, the free density and current are controlled from outside and correspond to the bare external sources, $\rhof=\nabla\cdot\DD=-\bar\rho$ (note how this absorbs the minus that appeared before) and $\jjf=\nabla\times\HH=\bAA^*$.

Thus the external sources adapt to the feedback from the quantum system just following the free Maxwell equations, so that the physically present fields $(\EE,\BB)$ come from a combination of the (freely adjusted) charge and current $(\rhof,\jjf)$ plus the charges and current due to the (bound) quantum system $(\rhob,\jjb)$. Overall we get
\begin{align}
    \rhof+\rhob &= -\bar\rho + \rho = \phi^*\quad\text{and} \\
    \jjf+\jjb &= \bAA^*-\AA^* = \jj,
\end{align}
which are exactly the parameters that enter the linear Pauli--Fierz Hamiltonian of Eq.~\eqref{eq:H}. Consequently, we effectively take a back-reaction on the environment into account when we switch to a Maxwell-regularized description, just as in a macroscopic Maxwell approach. The mixed quantities $(\bar\rho,\bAA)$ (or rather $(-\bar\rho,\bAA)$ to have the correct sign) can thus be seen as relating to an environment where no quantum system is present. And indeed with $\rho=0$ and $\AA=0$ we would simply have $\phi=\rhof^*=-\bar\rho^*$ and $\jj=\jjf=\bAA^*$. Through the coupling, the quantum system then acts back on the external sources which leads to the total values $(\phi,\jj)$ that finally enter our usual description.
Allowing the classical sources to react to the quantum system, instead of treating the environment as fixed, thus automatically leads to the Maxwell-regularization of the whole theory. In the present case this reaction is purely electrodynamical, but in principle more complex models for the environment are thinkable. Whenever the model provides the mathematical structure of Moreau--Yosida regularization and once we have reached self-consistency, we then always recover the results from the linear quantum theory.

A prime insight for the methodology of \ac{MaxQEDFT} comes from the fact that the basic quantities $(\bar\rho,\bAA)$ for the Maxwell-regularized universal functional are \emph{not} the internal density and vector potential that connect to the polarization and magnetization of the quantum medium by Eq.~\eqref{eq:MacroMax-P-M}. Instead the main variables of the new theory are the mixed quantities that correspond to the displacement and magnetizing fields $\DD$ and $\HH$ by Eq.~\eqref{eq:MacroMax-D-H}.
Indeed, we could rewrite all the functionals entirely in terms of the gauge-invariant fields from macroscopic Maxwell theory by substituting $(\phi,\jj)\to(\EE,\BB)$, $(\rho,\AA)\to(\PP,\MM)$, and $(\bar\rho,\bAA)\to(\DD,\HH)$.
The Fréchet derivative of the Maxwell-regularized universal functional $\rmd\bar F^\Lambda$ from Eq.~\eqref{eq:nabla-F} then defines a mapping $(\DD,\HH) \mapsto (\EE,\BB)$. In the context of the macroscopic Maxwell equations this is exactly the inversion of the \emph{constitutive relations}.
To make this explicit, the functional derivative $\rmd\bar F^\Lambda$ needs to be separated into its electric and magnetic components, $-\phi = \delta\bar F^\Lambda/\delta\bar\rho$ and $\jj = \delta\bar F^\Lambda/\delta\bAA$. This yields
\begin{align}
    \label{eq:const-rel-E}
    \EE &= -\nabla\phi = \nabla \left( \frac{\delta\bar F^\Lambda}{\delta\bar\rho} \right),\\
    \label{eq:const-rel-B}
    \BB &= \nabla\times\jj^* = \nabla\times \left( \frac{\delta\bar F^\Lambda}{\delta\bAA} \right)^*,
\end{align}
where the functional depends on $\bar\rho=-\nabla\cdot\DD$ and $\bAA = (\nabla\times\HH)^*$.
We stress that this allows to derive constitutive relations from the underlying quantum system to all orders in a rigorous way, while the unregularized functional $F^\Lambda(\rho,\AA)$ is known to be non-differentiable in the usual functional setting.
In order to perform the conventional linear expansion of this relation, we still have to assume existence of the second-order functional derivative that Moreau--Yosida regularization does not automatically guarantee. Then we can write for any chosen reference point $(\bar\rho_0,\bAA_0)\in\X$ that
\begin{widetext}
\begin{equation}\label{eq:dF-expansion}
    \frac{\delta \bar F^\Lambda}{\delta \bar\rho(\rr)} = \left. \frac{\delta \bar F^\Lambda}{\delta \bar\rho(\rr)} \right|_{(\bar\rho_0, \bAA_0)} \!\! + \int \!\! \underbrace{ \left. \frac{\delta^2 \bar F^\Lambda}{\delta \bar\rho(\rr) \delta\bar\rho(\rr')}\right|_{(\bar\rho_0, \bAA_0)} }_{\textstyle(\bar\chi^\Lambda_{\rme\rme})^{-1}(\rr,\rr')} \!\!\! ( \bar\rho(\rr')-\bar\rho_0(\rr') ) \,\rmd\rr' + \int \! \underbrace{ \left.\frac{\delta^2 \bar F^\Lambda}{\delta \bar\rho(\rr) \delta \bAA(\rr')}\right|_{(\bar\rho_0, \bAA_0)} }_{\textstyle(\bar\chi^\Lambda_{\rme\rmm})^{-1}(\rr,\rr')} \!\!\! (\bAA(\rr')-\bAA_0(\rr')) \,\rmd\rr' + \ldots
\end{equation}
\end{widetext}
The first integral kernel was already considered by \citet{Hohenberg1964} for the unregularized universal functional of DFT and then gives the usual inverse density-response function, while here the response is in terms of a mixed density (bare source). From Eq.~\eqref{eq:const-rel-E} we can further read off that it is related to the integral kernel of the inverse electric permittivity, since it gives the linearized connection between $\DD$ and $\EE$,
\begin{equation}
    \eps^{-1}_\rme(\rr,\rr') = \nabla (\bar\chi^\Lambda_{\rme\rme})^{-1} \otimes \nabla',
\end{equation}
where we employed the dyadic product.
The second integral kernel in Eq.~\eqref{eq:dF-expansion}, $(\bar\chi^\Lambda_{\rme\rmm})^{-1}$,  expresses the inverse of the $\bAA$ response from a potential variation, $\eps^{-1}_\rmm(\rr,\rr')$, discussed in Ref.~\cite{flick2019light} for the usual time-dependent case. A relation for the magnetic permeability can be derived in an analogous manner.

If, in contrast, we put the square-norm energy functional $\frac{1}{2}\|(\bar\rho,\bAA)\|^2_\X$ of just the bare electromagnetic field into the constitutive relations of Eqs.~\eqref{eq:const-rel-E}-\eqref{eq:const-rel-B} instead of $\bar F^\Lambda(\bar\rho,\bAA)$, the functional derivative evaluates like in Eq.~\eqref{eq:square-deriv-X} and we are left with
\begin{align}
    \EE &= \nabla\bar\rho^* = \eps_0^{-1}\DD, \\
    \BB &= \nabla\times\bAA = \mu_0\HH,
\end{align}
exactly the constitutive relations for the vacuum case.

In the physical picture developed in this section, the $\rmd \bar F^\Lambda$ thus directly relates to the electric permittivity (dielectric function) and magnetic permeability and further to the electric and magnetic susceptibility as the internal response of the quantum system that yields $\PP$ and $\MM$.
This clarifies how \ac{MaxQEDFT} reformulates the theory in terms of the physically relevant fields. The same treatment about relating a bare and a screened potential with a dielectric function appears in \citet[Sec.~5.2.1]{GiulianiVignale2005}.

Another important physical aspect that can be addressed with this functional-theoretic reformulation of non-relativistic \ac{QED} is the appearance of coupled light-matter equilibrium states that have recently received special attention and were termed \emph{endyonic} states~\cite{endyon2025}. In contrast to polaritons, which are light-matter hybrids that arise due to mixing of light and matter \emph{excitations}, endyons are \emph{hybrid equilibrium states}, i.e., they describe the possible coupled light-matter ground states that Pauli--Fierz theory is designed to predict. Such endyonic states have already been investigated with different ab initio \ac{QED} computational methods and at various levels of approximation~\cite{ruggenthaler2023understanding,mandal2023theoretical,bauman2025perspective}. Of specific interest are those collective light-matter ground states, where a large number of charged particles interact with the photon field and generate novel equilibrium states of matter~\cite{ruggenthaler2023understanding,sidler2025collectively}. The main obstacle for an unambiguous characterization of such states is, however, that already a single charged particle in the usual \ac{QED} approaches is a hybrid light-matter system that depends on the chosen cutoff $\Lambda$. It remains unknown whether or how fully renormalized light-matter systems emerge from a Pauli--Fierz description, which would be a prerequisite for a detailed characterization. Hitherto, in most cases, the standard assumption is that without any specifically designed photonic structure, e.g., optical cavities, the coupled light-matter ground state can be approximated well by the usual Schr\"odinger ground state and the photonic part only shows up through switching from bare to observable masses of the \emph{individual} charged elementary particles~\cite{ruggenthaler2023understanding}. Also, usually only the \emph{difference} to free-space modes is actively modeled~\cite{svendsen2025effective,eckhardt2025surface}. However, since a cavity itself is a physical system in free space, and nothing distinguishes fundamentally a system inside the cavity from the cavity itself, this argument becomes ultimately circular. Thus, in order to understand the different possible coupled vacua, we need to be able to remove the cutoff $\Lambda$ and really investigate the difference to situations that rely on the Schr\"odinger equation alone. It is obvious that there are situations where a simple Schr\"odinger description is inadequate and to identify those situations and to provide new computational tools for them is the main goal of ab initio \ac{QED}. Hence, in the next step, we will leverage the functional-theoretic reformulation and develop the necessary tools to do so.

\section{Removal of cutoff and Kohn--Sham system}
\label{sec:renorm-ks}

Let us next consider what happens as $\Lambda \rightarrow \infty$, i.e., when considering all frequencies of the photon field equally. There are different ways of how to take this limit. In analogy to the usual free-space ($\phiconf = 0$) renormalization, we adapt $m^{\Lambda}$ such that the Mie frequency (see Sec.~\ref{sec:setting}) stays constant while we increase $\Lambda$. We adopt a many-particle renormalization scheme since the common single-particle renormalization schemes of perturbative \ac{QED}, e.g., fixing the dispersion of the single free particle, might no longer be adequate~\cite{welakuh2025nonperturbative}. Instead, a corresponding observable many-body effect should be used that tells us how to absorb the self-interaction into an effective mass parameter. In other words, since for a general many-body system we will not be able to identify individual free particles, single-particle renormalization will not guarantee that we have taken into account the self-interaction correctly. 
This switch to an \emph{effective} mass in order to subsume certain \ac{QED} effects already reminds of the Kohn--Sham scheme in \ac{DFT}, where the potential is adjusted such that it subsumes interaction effects, which itself is again an \ac{QED} effect as we have stressed before. We will here go one step further and use this technique to include \emph{all} \ac{QED} effects including interactions by construction of a corresponding Kohn--Sham system while simultaneously removing the cutoff.

While the removal of the cutoff is a difficult problem on the level of the Hamiltonian of Eq.~\eqref{eq:H}, we instead have the possibility to discuss it on the level of the universal functional $\bar F^\Lambda(\bar\rho, \bAA)$ that is the central object of our theory and encodes all equilibrium properties. Indeed, at this point the Kohn--Sham construction, which is commonly thought of as a mere computational trick, becomes a natural tool to handle this limit. At the same time, this gives a more fundamental physical meaning to the auxiliary non-interacting non-coupled system.

\begin{figure}
\centering
\begin{tikzpicture}[scale=1]
    \draw[->] (0,2.5) -- (4,2.5)
        node[right] {$\rhoref+\X$};
    \draw[->]
        (1.37,2.2) -- (1.37,6);
    \draw[-]
        (2.7,2.6) -- (2.7,2.4) node[below] {$(\bar\rho,\bAA)$};
    \draw[<-]
        (1.32,2.45) -- (1.17,2.3) node[below left=-4pt] {$(\rhoref,0)$};
    \draw (0,4) parabola bend (1,3) (3,6) node[right] {$\bar F^\Lambda$};
    \draw[dashed] (0,2.41) -- (3,3.91) node[below right,xshift=-.5em] {$\bar F^\Lambda(\rhoref,0)+\langle \rmd\bar F^\Lambda(\rhoref,0),\cdot-(\rhoref,0)\rangle$};
    \draw[<->] (2.7,3.77) --node[right] {$\Fren^\Lambda(\bar\rho,\bAA)$} (2.7,5.18);
\end{tikzpicture}
\caption{Definition of the renormalizable functional as the Vainberg--Bregman distance.}
\label{fig:bregman}
\end{figure}
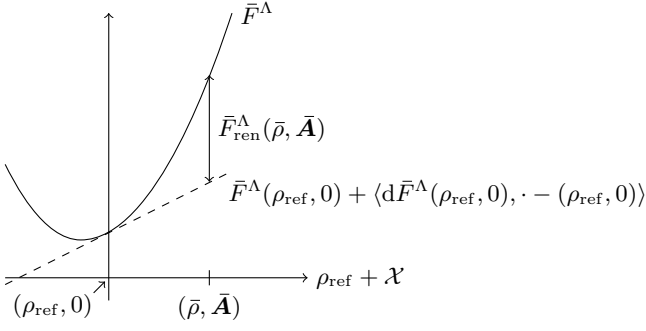

We start by defining a \emph{renormalizable} functional relative to the reference point $(\rhoref,0)$,
\begin{equation}\label{eq:def-Fren}\begin{aligned}
    \Fren^\Lambda(\bar\rho, \bAA) &= \bar F^\Lambda(\bar\rho,\bAA) - \bar F^\Lambda(\rhoref,0) \\
    &\quad- \langle \rmd \bar F^\Lambda(\rhoref,0),(\bar\rho-\rhoref, \bAA) \rangle.
\end{aligned}\end{equation}
In the zero-coupling limit ($\Lambda=0$) we get $\rmd \bar F^0(\rhoref,0) = (0,0)$ since the fixed reference point was defined for $\Lambda=0$ at $(\phi,\jj)=0$. This does not hold any more for $\Lambda>0$, i.e., systems with coupling.  Consequently, the renormalizable functional at $\Lambda=0$ reduces to $\Fren^0(\bar\rho, \bAA) = \bar F^0(\bar\rho,\bAA) - \bar F^0(\rhoref,0)$ and we obtain $\rmd\Fren^0(\bar\rho, \bAA) = \rmd\bar F^0(\bar\rho,\bAA)$.
The construction in Eq.~\eqref{eq:def-Fren} corresponds to what is called a \emph{Vainberg--Bregman distance}~\cite{Vainberg-book,RPK2026} and it is illustrated in Fig.~\eqref{fig:bregman}. It is ubiquitous in physics, specifically in quantum and classical thermodynamics, where a variant of it defines the \emph{relative entropy} (Kullback--Leibler divergence for probability distributions~\cite{Kullback-Leibler1951} and Umegaki relative entropy for quantum states~\cite{Umegaki1962}) that encodes the information-difference between two configurations. It also appears in a maximum-entropy derivation of thermal \ac{DFT}, where a reference is included in the same way to serve as prior information~\cite{Yousefi2023}. The possibility to compute quantum systems relative to a chosen reference, similar to adding the \emph{non-linear} entropy to standard \emph{linear} quantum physics, can be seen as the main advantage over a pure linear wave-function formulation. Since the renormalizable functional $\Fren^\Lambda$ from Eq.~\eqref{eq:def-Fren} is the difference between $\bar F^\Lambda$ and its tangential approximation at the reference point, it is again convex and differentiable. It is also directly seen to be non-negative with minimum $\Fren^\Lambda(\rhoref,0)=0$.
For any $\Lambda$ we let
\begin{equation}\label{eq:ks-interacting}
    (-\phi^\Lambda,\jj^\Lambda) = \rmd\Fren^\Lambda(\bar\rho, \bAA) = \rmd\bar F^\Lambda(\bar\rho,\bAA)-\rmd\bar F^\Lambda(\rhoref,0).
\end{equation}
From the Lipschitz-continuity of $\rmd \bar F^\Lambda$ for all $\Lambda$ (with constant 1 as noted after Eq.~\eqref{eq:F-regularized-alt}) it then directly follows that
\begin{equation}
    \|(\phi^\Lambda,\jj^\Lambda)\|_{\X^*} \leq \|(\bar\rho-\rhoref,\bAA)\|_\X.
\end{equation}
So for any fixed $(\bar\rho,\bAA)$ we have $(\phi^\Lambda,\jj^\Lambda)$ as a sequence with increasing $\Lambda$ bounded in the Hilbert space $\X^*$.
From integrating the derivative of $\Fren^\Lambda$ from the reference point to $(\bar\rho,\bAA)$, this allows us to derive the estimate
\begin{equation}\begin{aligned}
    0 &\leq \Fren^\Lambda(\bar\rho,\bAA) = \Fren^\Lambda(\rhoref,0) \\
    &+ \int_0^1 \langle \rmd\Fren^\Lambda(\rhoref+t(\bar\rho-\rhoref), t\bAA), (\bar\rho-\rhoref,\bAA) \rangle \rmd t\\
    &\leq \frac{1}{2}\|(\bar\rho-\rhoref,\bAA)\|_{\X}^2.
\end{aligned}\end{equation}
Consequently, the renormalizable functional is bounded from above and below. Yet, every bounded sequence has a convergent subsequence, and so we can write $\Fren^\Lambda(\bar\rho,\bAA) \to \Fren^\alpha(\bar\rho,\bAA)$, where we introduce $\alpha$ to enumerate the different possible pointwise limits. The resulting (non-unique) \emph{renormalized} functional defined this way is again convex, but not necessarily differentiable or lower semicontinuous. This fits to the idea that in the limit $\Lambda\to\infty$ one generally loses connection to a well-defined Hamiltonian and thus also to a microscopic quantum theory on the \emph{original} separable Hilbert space (since lower-semicontinuity is a prerequisite for that~\cite{MY-Perspective}). But everything remains well-defined on the level of the functional theory.

Also, for the $(\phi^\Lambda,\jj^\Lambda)$ as a bounded sequence in $\X^*$ we have a weakly convergent subsequence by the Banach--Alaoglu theorem for Hilbert spaces~\cite[Cor.~11.9]{clason2020-book}. Similar as above, we write $(\phi^\Lambda,\jj^\Lambda)\weakto (\phi^\alpha,\jj^\alpha)\in\X^*$ again.
Remember that weak convergence means that for the chosen subsequence it holds
\begin{equation}\label{eq:phiLambdaLimit}
    \langle (\phi^\Lambda,\jj^\Lambda),(f,\aa)\rangle \to \langle (\phi^\alpha,\jj^\alpha),(f,\aa) \rangle,   
\end{equation}
with finite limits for every $(f,\aa)\in\X$, so the potential energy for $(\phi^\alpha,\jj^\alpha)$ is given by the limits above. This yields $(-\phi^\alpha,\jj^\alpha) \in \subdiff\Fren^\alpha(\bar\rho,\bAA)$ from the subdifferential of the renormalized functional.
We note that this enumeration of different limit points with $\alpha$ reminds of order parameters introduced to describe phase transitions, like magnetization symmetry breaking in the Ising model during the thermodynamic limit, which could be handled in the same way~\cite{Haag1996LocalQuantumPhysics, thirring2002quantum}. At this point, we have successfully removed the cutoff within our equilibrium \ac{QED} functional theory and we always find sources $(\phi^\alpha,\jj^\alpha)$ connecting to internal $(\rho^\alpha,\AA^\alpha)=(\bar\rho+(\phi^\alpha)^*,\bAA-(\jj^\alpha)^*)$ that reproduce any choice of $(\bar\rho,\bAA)$ through the renormalized functional $\Fren^\alpha$. Yet, as in the case of the physics of phase transitions, there might not exist a well-defined interacting microscopic theory on the original Hilbert space. Looking back at Sec.~\ref{sec:interpret}, we note that the resolution of the renormalization dilemma of standard linear quantum physics is that we have now connected the bare sources $(-\bar \rho, \bAA)^*$ to their screened and renormalized counterparts $(\phi^\alpha,\jj^\alpha)$ and hence the renormalized quantum system just appears as modifying the external potential differently. To make this fully renormalized \ac{QED} theory practical, the trick is to employ the Kohn--Sham system, which does not suffer from any renormalization issues, and that gives the exact same results as the $\Lambda \rightarrow \infty$ theory by including a corrective renormalization potential and current on the original Hilbert space.

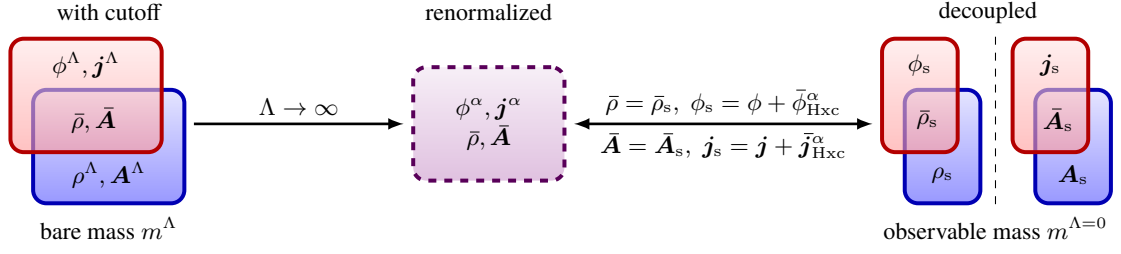
\begin{figure*}[ht]
\begin{tikzpicture}[
  box/.style={
    rectangle, rounded corners=5pt,
    draw=blue!70!black, line width=1.5pt,
    top color=blue!10, bottom color=blue!40,
    minimum width=2cm, minimum height=1.5cm,
    inner sep=5pt, align=center
  },
  boxalt/.style={
    rectangle, rounded corners=5pt,
    draw=red!70!black, line width=1.5pt,
    top color=red!10, bottom color=red!40,
    minimum width=2cm, minimum height=1.5cm,
    inner sep=5pt, align=center,
    fill opacity=0.6
  },
  boxalt2/.style={
    rectangle, rounded corners=5pt, dashed,
    draw=violet!70!black, line width=1.5pt,
    top color=violet!10, bottom color=violet!40,
    minimum width=2cm, minimum height=1.5cm,
    inner sep=5pt, align=center,
    fill opacity=0.6
  },
]

\node[box] (box51) {};
\node[boxalt] (box52) at (box51.north) [yshift=-0.1cm,xshift=-.3cm] {};
\node at (box52.north) [yshift=-0.37cm] {$\phi^\Lambda,\jj^\Lambda$};
\node at (box52.north) [yshift=-1.1cm,xshift=.1cm] {$\bar\rho, \bAA$};
\node at (box51.north) [yshift=-1.15cm] {$\rho^\Lambda,\AA^\Lambda$};
\node at (box51.north) [yshift=1cm] {with cutoff};
\node at (box51.south) [yshift=-.3cm] {bare mass $m^\Lambda$};
\node[boxalt2] (boxrenorm) at (box51.east) [yshift=.3cm,xshift=4cm] {};
\draw[-latex,line width=1pt] ([yshift=.3cm,xshift=.1cm]box51.east) -- node[midway,above] {$\Lambda\to\infty$} ([xshift=-.1cm]boxrenorm.west);
\node[align=center] at (boxrenorm.center) {$\phi^\alpha,\jj^\alpha$ \\ $\bar\rho, \bAA$};
\node at (boxrenorm.north) [yshift=0.7cm] {renormalized};
\node[box,minimum width=1cm] at (box51.east) [xshift=10cm] (box61) {};
\node[boxalt,minimum width=1cm] (box62) at (box61.north) [yshift=-0.1cm,xshift=-.3cm] {};
\node at (box62.north) [yshift=-0.37cm] {$\phis$};
\node at (box62.north) [yshift=-1.1cm,xshift=.1cm] {$\brhos$};
\node at (box61.north) [yshift=-1.15cm] {$\rhos$};
\node at (box61.north) [yshift=1cm,xshift=.6cm] {decoupled};
\node at (box61.south) [yshift=-.3cm,xshift=.7cm] {observable mass $m^{\Lambda=0}$};
\draw[latex-latex,line width=1pt] ([yshift=.3cm,xshift=-.4cm]box61.west) -- node[midway,above] {$\bar\rho=\brhos, \; \phis=\phi+\phiMregHxc^\alpha$} node[midway,below] {$\bAA=\bAAs, \; \jjs=\jj+\jjMregHxc^\alpha$} ([xshift=.1cm]boxrenorm.east);
\node[box,minimum width=1cm] at (box61.east) [xshift=1.2cm] (box71) {};
\node[boxalt,minimum width=1cm] (box72) at (box71.north) [yshift=-0.1cm,xshift=-.3cm] {};
\node at (box72.north) [yshift=-0.37cm] {$\jjs$};
\node at (box72.north) [yshift=-1.1cm,xshift=.1cm] {$\bAAs$};
\node at (box71.north) [yshift=-1.15cm] {$\AAs$};
\draw[dashed] ([xshift=-.2cm]box72.north west) -- ([xshift=-.5cm]box71.south west);

\end{tikzpicture}
\caption{Schematic for the removal of the cutoff on the external quantities and their further connection to a quantum system with the same mixed density $\brhos=\bar\rho$ from an effective potential $\phis$, as well as a decoupled mixed vector potential $\bAAs=\bAA$ from an external current $\jjs$.}
\label{fig:KS-renorm}
\end{figure*}

We define the canonical Kohn--Sham system by setting $\Lambda=0$ and thus removing any coupling to the photon field. This also removes the interaction between the particles, such that they are only affected by an effective potential $\phiconf+\phis$, where we keep the same offset and choose a $\phis\in\potSpace$. We take the ground-state density of this Hamiltonian to be $\rhos$ with $\brhos = \rhos - \phis^*\in\densSpace$.
Since the photon field also does not couple to any external sources, we simply have $\AAs=0$ from the expectation value of the vector-potential operator and thus $\bAAs=\jjs^*$.
Together, the mixed quantities $(\rhos,\AAs)\in\X$ then describe the Kohn--Sham system where one chooses $(\phi_s,\jj_s)\in\X^*$ in such a way that the mixed quantities of the interacting system are reproduced,
\begin{equation}\label{eq:rho-rhos}
    (\brhos,\bAAs) = (\bar\rho,\bAA).
\end{equation}
We see that for the field part this is trivially possible by setting $\jjs=\bAA^*$. But also for the particle part we always find a solution via
\begin{equation}\label{eq:ks-noninteracting}
    (-\phis,\jjs) = \rmd\bar F^0(\bar\rho,\bAA) = \rmd\Fren^0(\bar\rho,\bAA).
\end{equation}
This uses the universal functional for the non-interacting non-coupled system ($\Lambda=0$), which includes the trivial solution for $\jjs$ in the notation. Remember that from what was said before, it does not matter if we take the functional $\bar F^0$ or its renormalizable version here. Fig.~\ref{fig:KS-renorm} displays the whole procedure schematically.
We can now combine Eqs.~\eqref{eq:ks-interacting} and \eqref{eq:ks-noninteracting} to define the \emph{Maxwell-regularized \ac{Hxc} potential and current} in full analogy to \ac{DFT},
\begin{equation}\label{eq:ren-pot}\begin{aligned}
    (\phiMregHxc^\Lambda[\bar\rho,\bAA],-\jjMregHxc^\Lambda[\bar\rho,\bAA]) &= \rmd\!\left(\Fren^\Lambda - \bar F^0 \right)(\bar\rho,\bAA) \\
    &= (-\phi^\Lambda,\jj^\Lambda) - (-\phis,\jjs).
\end{aligned}\end{equation}
The resulting $(\phiMregHxc^\Lambda,\jjMregHxc^\Lambda)$ can be combined with $(\phi^\Lambda,\jj^\Lambda)$ in order to exactly reproduce the same $(\bar\rho,\bAA)$ with a non-interacting non-coupled Kohn--Sham system. Following the equation above, this requires $\phis = \phi^\Lambda + \phiMregHxc^\Lambda$ and $\jjs=\jj^\Lambda + \jjMregHxc^\Lambda$ in order to correct for the \ac{QED} effects. 
The Maxwell-regularized \ac{Hxc} potential and current work in the same way as the counterterm in mass renormalization, which is the difference between the mass at $\Lambda>0$ (bare mass) and $\Lambda=0$ (renormalized mass) and can thus be equally seen as a `renormalization potential (and current)'.

We now see that the limit from before carries over to $\phis = \phi^\alpha + \phiMregHxc^\alpha$ and $\jjs=\jj^\alpha + \jjMregHxc^\alpha$ and we can define $(\phiMregHxc^\alpha[\bar\rho,\bAA],\jjMregHxc^\alpha[\bar\rho,\bAA])\in\X^*$ for every $(\bar\rho,\bAA)\in\X$.
This provides an elegant tool to resolve renormalization issues purely on the level of functional theory, where in the Kohn--Sham system we have $\Lambda=0$ and no issues arise. We will highlight later (in Sec.~\ref{sec:discussion}) that a similar procedure is already implicitly used in \ac{DFT} and \ac{QEDFT} regularly.
We have thus succeeded in removing the cutoff within our theory and have obtained a finite but non-linear effective theory for \ac{QED} by the Kohn--Sham construction. While there is no bare particle picture in \ac{QED}, this theory allows for a standard quantum-mechanical description of non-interacting particles within an effective potential. To get a ground-state solution, the problem can be solved iteratively by employing the Maxwell-regularized \ac{Hxc} potential and current self-consistently. Choose any $(\phi,\jj)\in\X^*$ for the interacting system (for the full $\Lambda\to\infty$ theory, so this also means choosing an order parameter $\alpha$ in case the limit procedure described before is not unique), then solve iteratively starting with a guess $(\bar\rho_0,\bAA_0)$,
\begin{equation}\label{eq:ks-step}
    (\bar\rho_{i+1},-\bAA_{i+1}) \in \supdiff\bar E^0(\phi+\phiMregHxc^\alpha[\bar\rho_i,\bAA_i],\jj+\jjMregHxc^\alpha[\bar\rho_i,\bAA_i]).
\end{equation}
The evaluation of this formula splits into two separate problems since particles and field decouple in the Kohn--Sham system. In every step we can independently solve for the next vector potential with the Maxwell duality
\begin{equation}
    \bAA_{i+1} = (\jj+\jjMregHxc^\alpha[\bar\rho_i,\bAA_i])^*
\end{equation}
as described before and get $\bar\rho_{i+1}$ from
\begin{equation}
    \bar\rho_{i+1} = \rho_{i+1} - (\phi+\phiMregHxc^\alpha[\bar\rho_i,\bAA_i])^*,
\end{equation}
where $\rho_{i+1}$ is the ground-state solution of the non-interacting Schr\"{o}dinger equation with external potential $\phi+\phiMregHxc^\alpha[\bar\rho_i,\bAA_i]$.
Note that despite having decoupled particle and field parts, we still rely on the full density \emph{and} vector potential information $(\bar\rho_i,\bAA_i)$ in every step when evaluating the Maxwell-regularized \ac{Hxc} potential and current.
Since this procedure can lead to degeneracy even after choosing a specific $\alpha$, the solution is not necessarily unique. This is also why in Eq.~\eqref{eq:ks-step} the next step is the element of a superdifferential, possibly including multiple densities. 
It is important to note that this formulation of the Kohn--Sham method in the \ac{MaxQEDFT} setting does \emph{not} aim at the usual density to agree between the interacting and non-interacting systems, but instead at $(\bar\rho,\bar\AA)$ pairs.
This also means that the orbitals of the Kohn--Sham method described here will necessarily be different to those from the usual method. Indeed, from Sec.~\ref{sec:interpret} we learned that we are instead actually making the bare sources agree. Consequently, in \ac{MaxQEDFT}, we put a renormalized interacting system and a non-interacting non-coupled system into exactly the same bare environment $(-\bar\rho,\bAA)^*$ on which they then exert different back-reactions to yield effective sources $(\phi^\alpha,\jj^\alpha)$ and $(\phis,\jjs)$. As we will see in Sec.~\ref{sec:MaxDFT}, the same can be done in a Maxwell-regularized version of standard \ac{DFT}, and thus we can finally and unambiguously compare the difference between a purely quantum-mechanical ground state and of a fully coupled and renormalized \emph{endyonic} state of light and matter.

The Kohn--Sham method adapted to Moreau--Yosida regularized functionals was first described by \citet{Kvaal2014} in a Hilbert-space setting and by \citet{KSpaper2018} for more general Banach spaces.
Convergence of the iteration procedure was proved with an additional damping step in \emph{finite} dimensions~\cite{KS_PRL_2019,KS_PRL_2019_Erratum,MY-Perspective}. We expect that this convergence result can be achieved similarly in the \ac{MaxQEDFT} setting, where the confining potential $\phiconf$ yields the necessary compactness result.

\section{Maxwell-regularized DFT}
\label{sec:MaxDFT}

In this and the following section, we discuss two boundary cases that remove the quantized photon field from the description. In the first one we remove $\AA$ entirely from the description, but still include \ac{QED} effects by switching from the bare mass $m^\Lambda$ to the renormalized (i.e., the observable) mass $m$ and by using the singular Coulomb interaction as in Eq.~\eqref{eq:interact-limit-Coulomb} (where no cutoff is necessary).
The resulting Hamiltonian is then the standard electron-only Hamiltonian of \ac{QM} and the ensuing theory has the same descriptive power as \ac{DFT}. The universal functional $F(\rho)$ is replaced by its Maxwell-regularized version like in Eq.~\eqref{eq:F-regularized} that is the total internal energy of the quantum system and the electrostatic field.
We thus have $\bar\rho = \rho - \phi^*$, where $\rho$ is the ground-state density for potential $\phi$, and
\begin{equation}
    \bar F(\bar\rho) = F(\rho) + \frac{1}{2}\|\phi\|_\potSpace^2 = \inf_{\rho'}\left\{ F(\rho') + \frac{1}{2}\|\bar\rho-\rho'\|_{\densSpace}^2 \right\}.
\end{equation}
By exactly the same arguments as in Sec.~\ref{sec:df-formulation}, the functional $\bar F(\bar\rho)$ becomes differentiable and also all other results hold in the same way by just ignoring the vector-potentials $\AA$ and $\bAA$.
In particular, by setting up an equally differentiable non-interacting Maxwell-regularized functional $\bar F^0$ representing the Kohn--Sham system, we are able to define a Maxwell-regularized version of the \ac{Hxc} energy functional as $\EMregHxc(\bar\rho) = \bar F(\bar\rho)-\bar F^0(\bar\rho)$. As in Eq.~\eqref{eq:ren-pot}, we then get the Maxwell-regularized \ac{Hxc} potential in a well-defined manner as its gradient,
\begin{equation}\label{eq:phiren}\begin{aligned}
    \phiMregHxc[\bar\rho] &= \rmd \bar F(\bar\rho)-\rmd \bar F^0(\bar\rho) \\&= \rmd (\bar F-\bar F^0)(\bar\rho) = \rmd \EMregHxc(\bar\rho),
\end{aligned}\end{equation}
which holds since taking the functional derivative is a linear operation.
As already noted elsewhere~\cite[Sec.~VIII]{Penz2023-PartI}, this stands in contrast to standard \ac{DFT}, where the same does not rigorously hold for the usual \ac{Hxc} energy functional $\EHxc(\rho)=F(\rho)-F^0(\rho)$, since the subdifferential has to be employed instead that does not obey this linearity,
\begin{equation}
    \phiHxc[\rho] \in \subdiff F(\rho)-\subdiff F^0(\rho) \neq \subdiff (F-F^0)(\rho) = \subdiff \EHxc(\rho).
\end{equation}
Even more problematically, the subdifferential $\subdiff F(\rho)$ is an empty set whenever the density $\rho$ is not $v$-representable by the ground-state of an interacting system. Similarly, $\subdiff F^0(\rho)$ is empty if $\rho$ is not non-interacting $v$-representable. Formulated in terms of the Maxwell-regularized functional with mixed quantities, this issue is completely avoided. This can be seen as a major advancement from standard \ac{DFT} and we propose to call the resulting theory \ac{MaxDFT}. We next discuss some important properties of the regularized functional for linking to a Kohn--Sham system in this setting that have partly been described before~\cite{MY-Perspective}.

Remember that the Kohn--Sham system in \ac{MaxDFT} is set up by demanding agreement of the mixed densities between an interacting and a non-interacting system as before in Eq.~\eqref{eq:rho-rhos}, instead of demanding this from the internal densities as in standard \ac{DFT}.
Taking any mixed density $\bar\rho=\brhos$, we have potentials $\phi=-\rmd\bar F(\bar\rho)$ for the interacting system and $\phis=-\rmd \bar F^0(\brhos)$ for an auxiliary Kohn--Sham system. This gives $\rho=\bar\rho+\phi^*$ and $\rhos=\brhos+\phis^*$ and we arrive at
\begin{equation}
    \bar\rho = \rho-\phi^* \overset{!}{=} \brhos = \rhos-\phis^* = \rhos - (\phi^* + \phiMregHxc^*[\bar\rho])
\end{equation}
from our central relation Eq.~\eqref{eq:def-mixed} and the definition of the Maxwell-regularized \ac{Hxc} potential in Eq.~\eqref{eq:phiren}. This means that
\begin{equation}\label{eq:Hxc-density}
    \phiMregHxc[\bar\rho] = \phis-\phi = (\rhos-\rho)^*
\end{equation}
and suggests to define a \ac{Hxc} charge density $\rhoHxc = \rhos-\rho$ that relates back to the Maxwell-regularized \ac{Hxc} simply through Maxwell-duality. As the difference between two internal densities we have $\rhoHxc\in\densSpace\cap L^1(\R^3)$ with a zero total charge (since $\rho$ and $\rhos$ both integrate to the same number of particles). Thus, $\phiMregHxc$ must decay exponentially at large distances as the solution to the Poisson equation for a density distribution $\rhoHxc$ with zero total charge. This is in contrast to the usual \ac{Hxc} potential that decays as $\sim|\rr|^{-1}$.
Such densities have been considered before as exchange-correlation charges for the standard Kohn--Sham construction~\cite{Gorling1999,LiuAyersParr1999,Callow2020} and should not be confused with the exchange-correlation hole.

The most important quantity of a Kohn--Sham calculation is arguably the ground-state energy, so we also need a way how to connect back to $E(\phi)$ of the interacting system from an evaluation of the Kohn--Sham ground state in \ac{MaxDFT}.
At the solution point $\bar\rho$ for a given potential $\phi$ the Maxwell-regularized Kohn--Sham energy $\bar E^0$ and interacting energy $\bar E$ are connected through
\begin{equation}\begin{aligned}
    \bar E(\phi)&=\bar F(\bar\rho)+\langle \phi,\bar\rho \rangle \\
    &=\bar F^0(\bar \rho) + \bar F(\bar\rho) - \bar F^0(\bar\rho) +\langle \phi -\phis+\phis,\bar\rho \rangle \\
    &=\bar E^0(\phis) + \EMregHxc(\bar\rho) -\langle \phiMregHxc[\bar\rho],\bar\rho \rangle,
\end{aligned}\end{equation}
where $\phis = \phi+\phiMregHxc[\rho]$ and $\bar\rho=\brhos$. The basic relation to the original energies without taking the external field energy into account is given by Eq.~\eqref{eq:regularizedE},
\begin{equation}
    \bar E(\phi) = E(\phi)-\frac{1}{2}\|\phi\|^2_\potSpace, \quad
    \bar E^0(\phis) = E^0(\phis)-\frac{1}{2}\|\phis\|^2_\potSpace.
\end{equation}
This combines to the energy expression for the usual ground-state energy,
\begin{equation}\begin{aligned}
    E(\phi)&=E^0(\phis)+\EMregHxc(\bar\rho)-\langle\phiMregHxc[\bar\rho],\bar\rho\rangle\\
    &+\frac{1}{2}\|\phi\|^2_\potSpace-\frac{1}{2}\|\phis\|^2_\potSpace.
\end{aligned}\end{equation}

We can now wonder how the huge arsenal of highly-developed \ac{Hxc} approximations of standard \ac{DFT} can be brought to use within the new theory. Indeed, the relation in Eq.~\eqref{eq:Hxc-density} can be employed to translate any approximation for ground-state solutions of the interacting system into a Maxwell-regularized \ac{Hxc} functional. In $\phiMregHxc^*=\rhos-\rho$, we let $\rhos=\supdiff E^0(\phis)$ and $\rho=\supdiff E(\phi)$, where for simplicity we assume a unique ground-state density in both systems. In case of degeneracy, the densities must be chosen from the superdifferentials such that they yield a matching $\bar\rho=\brhos$. Then, with $\phis=\phi + \phiMregHxc$, we arrive at the recursive relation
\begin{equation}
    \phiMregHxc^* = \supdiff E^0(\phi + \phiMregHxc) - \supdiff E(\phi).
\end{equation}
By using any approximation for the interacting system in $\supdiff E(\phi)$, we can use this scheme to self-consistently solve for a local potential (note how in this respect this procedure resembles the \ac{OEP} method~\cite{Kuemmel2008}). For example, we can put in a standard \ac{DFT} approximation represented by a \ac{Hxc} potential $\phiHxc$ of the form $\supdiff E(\phi) \approx \supdiff E^0(\phi + \phiHxc)$ to convert it into its Maxwell-regularized form.

In a very similar way, a Dyson-like equation can be derived for the Maxwell-regularized functional~\cite[Sec.~IV.C]{MY-Perspective}. The starting point is again the fundamental relation, now written as $-\phi=\bar\rho^*-\rho^*$, that can be used directly to set up a recursive scheme. For this, just choose $\rho\in\supdiff E(\phi)$ and put in $\phi=-\rmd\bar F(\bar\rho)$ to get
\begin{equation}
    \rmd\bar F(\bar\rho) \in \bar\rho^* - \supdiff E(-\rmd\bar F(\bar\rho))^*.
\end{equation}
If we just use the Hartree approximation $\supdiff E(\phi)\approx\rmd E^{\rm H}(\phi)$ (note that it is differentiable by the relation $E^{\rm H}(-\bar\rho^*)=\bar F^0(\bar\rho)-\frac 1 2 \|\bar\rho\|_{\densSpace}^2$ derived after Eq.~\eqref{eq:Fbar-variational} and thus yields a unique result) we arrive exactly at a form of the \ac{RPA}~\cite[Sec.~5.3.1]{GiulianiVignale2005}, where the notions of bare and screened potential as discussed in Sec.~\ref{sec:interpret} appear again,
\begin{equation}
    \rmd\bar F^{\rm RPA}(\bar\rho) = \bar\rho^* - \rmd E^{\rm H}(-\rmd\bar F^{\rm RPA}(\bar\rho))^*.
\end{equation}

\section{Mean-field vector-potential formulation}
\label{sec:MaxMDFT}

The second simplification does not remove the photon field components altogether, but replaces them with their expectation value with respect to a trial state in the form of a mean-field approximation, $\hAA^\Lambda \to \AA=\langle\hAA^\Lambda\rangle$, where we keep the cutoff but, as before, do not indicate it explicitly for the expectation value. The resulting Hamiltonian is
\begin{equation}\label{eq:Ham-mf}
\begin{aligned}
\hat{H}^{\Lambda,\mf}_{\phi,\jj} &= \sum_{i=1}^{N} \frac{1}{2m^\Lambda}\left[\bm{\sigma}_i\cdot \left(-\rmi\hbar\nabla_i + e\AA(\rr_i) \right)  \right]^2 \\
& + \sum_{i<j}^{N} e^2\wC^\Lambda(\rr_i-\rr_j) + \frac{1}{2\mu_0}\int |\nabla\times\AA(\rr)|^2\rmd\rr
\\
& - \sum_{i=1}^{N}e(\phiconf(\rr_i) + \phi(\rr_i)) - \int \AA(\rr) \cdot \jj(\rr) \rmd \rr. 
\end{aligned}
\end{equation}
We note that this is the usual electronic Hamiltonian with coupling to a vector potential and thus allows to consider magnetic fields, but it also still includes the magnetic-field energy and the coupling between vector potential and external current as scalars. We expand the square term by applying the identity
\begin{equation}
    (\bm{\sigma}\cdot\bm{u})(\bm{\sigma}\cdot\bm{v}) = \bm{u}\cdot\bm{v} + \rmi \bm{\sigma}\cdot(\bm{u}\times\bm{v})
\end{equation}
with
$\bm{u}=\bm{v}=-\rmi\hbar\nabla + e\AA$.
This gives
\begin{equation}
    (-\rmi\hbar\nabla + e\AA)^2 = -\hbar^2\Delta -\rmi\hbar e\AA\cdot\nabla-\rmi\hbar e\nabla\cdot\AA + e^2|\AA|^2
\end{equation}
and
\begin{equation}
   (-\rmi\hbar\nabla + e\AA)\times(-\rmi\hbar\nabla + e\AA) = -\rmi\hbar e (\nabla\times\AA).
\end{equation}
Expanding the cross product, all terms but one drop and we are left with the magnetic field.
In order to write this in a more concise form, we introduce the paramagnetic current operator with charge $-e$,
\begin{equation}
    \hJJ^\mathrm{p}(\rr) = -\frac{\hbar e}{2\rmi  m^\Lambda}\sum_{i=1}^{N}(\nabla_i\delta(\rr-\rr_i)+\delta(\rr-\rr_i)\nabla_i),
\end{equation}
and the spin magnetic moment density operator, also with charge $-e$,
\begin{equation}
    \hat\mm(\rr) = -\frac{\hbar e}{2m^\Lambda}\sum_{i=1}^{N}\bm{\sigma}_i\delta(\rr-\rr_i).
\end{equation}
Then the mean-field Hamiltonian can be rewritten as
\begin{equation}\label{eq:Ham-mf2}\begin{aligned}
    &\hat{H}^{\Lambda,\mf}_{\phi,\jj} = -\frac{\hbar^2}{2m^\Lambda}\sum_{i=1}^{N}\nabla_i^2 + \sum_{i<j}^{N} e^2\wC^\Lambda(\rr_i-\rr_j) - \sum_{i=1}^{N}e\phiconf(\rr_i) \\[-0.5em]
    & - \sum_{i=1}^{N}e \phi(\rr_i) -\! \int \!\!\AA(\rr)\!\cdot\!\left(\frac{e}{2m^\Lambda}\AA(\rr)\hat\rho(\rr) + \jj(\rr)+ \hJJ^\mathrm{p}(\rr)\right)\!\rmd\rr \\
    &+ \int (\nabla\times\AA(\rr)) \cdot \left( -\hat\mm(\rr)+ \frac{1}{2\mu_0} \nabla\times\AA(\rr) \right) \rmd\rr.
\end{aligned}\end{equation}
The first three terms are collected as the internal Hamiltonian $\hat H_{\rm{int}}^\Lambda$ (including $\phiconf$ and a cutoff $\Lambda$), while all other terms describe the coupling to the potential $\phi$, to $\AA$ and its magnetic field, and the energy of the magnetic field. 
We now take the expectation value of this Hamiltonian and write $\JJ^\mathrm{p}(\rr)=\langle\hJJ^\mathrm{p}(\rr)\rangle$ and $\mm(\rr)=\langle\hat\mm(\rr)\rangle$ for brevity. 
Using integration by parts and the vector-Laplacian identity together with Coulomb gauge ($\nabla\cdot\AA=0$) like before yields
\begin{equation}\label{eq:Hmf-expval}\begin{aligned}
    \langle\hat{H}_{\phi,\jj}^{\Lambda,\mf}&\rangle = \langle\hat H_{\rm{int}}^\Lambda\rangle + \langle\phi,\rho\rangle - \int \AA(\rr)\cdot\bigg( \frac{e}{2m^\Lambda}\AA(\rr)\rho(\rr) \\
    &+ \jj(\rr)+ \JJ^\mathrm{p}(\rr) +\nabla\times\mm(\rr) + \frac{1}{2\mu_0} \Delta_\mathrm{vec}\AA(\rr) \bigg) \rmd\rr
\end{aligned}\end{equation}
Not forgetting that, due to the negative charge, it holds $\rho \leq 0$ and that $-\Delta_\mathrm{vec}$ is a positive operator, we note that this expression is convex in $\AA$ and we can thus find a minimum by explicit component-wise differentiation w.r.t.\ $A_k$. This directly gives 
\begin{equation}\label{eq:MDFT-Maxwell}
    - \frac{1}{\mu_0} \Delta_\mathrm{vec}\AA =\jj+\JJ^\mathrm{p}+\frac{e}{m^\Lambda}\AA\rho+\nabla\times\mm,
\end{equation}
where we recognize that the charged diamagnetic current $\JJ^\mathrm{d} = (e/m^\Lambda)\AA\rho$ combines with the paramagnetic current and the spin current to form the total (physical) current $\JJ=\JJ^p+\JJ^d+\nabla\times\mm$. Equation~\eqref{eq:MDFT-Maxwell} is thus the static Amp\`{e}re--Maxwell law from Eq.~\eqref{eq:AmpereMaxwell} in Coulomb gauge that gets coupled to the usual Schr\"{o}dinger equation with vector potential.

\begin{figure}
\resizebox{.85\columnwidth}{!}{%
\begin{tikzpicture}[
  box/.style={
    rectangle, rounded corners=5pt,
    draw=blue!70!black, line width=1.5pt,
    top color=blue!10, bottom color=blue!40,
    minimum width=2cm, minimum height=1.5cm,
    inner sep=5pt, align=center
  },
  boxalt/.style={
    rectangle, rounded corners=5pt,
    draw=red!70!black, line width=1.5pt,
    top color=red!10, bottom color=red!40,
    minimum width=2cm, minimum height=1.5cm,
    inner sep=5pt, align=center,
    fill opacity=0.6
  },
]

\node[box] (box51) {};
\node[boxalt] (box52) at (box51.north) [yshift=-0.1cm,xshift=-.3cm] {};
\node at (box52.north) [yshift=-0.37cm] {$\phi,\jj$};
\node at (box52.north) [yshift=-1.1cm,xshift=.1cm] {$\bar\rho, \bAA$};
\node at (box51.north) [yshift=-1.15cm] {$\rho,\AA$};
\node at (box51.north) [yshift=1cm] {MaxQEDFT};

\node[box,minimum width=1cm] at (box51.east) [xshift=3cm,yshift=1.7cm] (box61) {};
\node[boxalt,minimum width=1cm] (box62) at (box61.north) [yshift=-0.1cm,xshift=-.3cm] {};
\node at (box62.north) [yshift=-0.37cm] {$\phi$};
\node at (box62.north) [yshift=-1.1cm,xshift=.1cm] {$\bar\rho$};
\node at (box61.north) [yshift=-1.15cm] {$\rho$};
\node at (box61.north) [yshift=1cm] {MaxDFT};
\draw[-latex,line width=1pt] ([yshift=.35cm,xshift=.1cm]box51.east) -- ([yshift=.35cm,xshift=-.4cm]box61.west) node[above left,align=center] {$\Lambda\to\infty$ \\ no $\hAA^\Lambda$} ;

\node[box,minimum width=1cm] at (box51.east) [xshift=3cm,yshift=-1.7cm] (box71) {};
\node[boxalt,minimum width=1cm] (box72) at (box71.north) [yshift=-0.1cm,xshift=-.3cm] {};
\node at (box72.north) [yshift=-0.37cm] {$\phi$};
\node at (box72.north) [yshift=-1.1cm,xshift=.1cm] {$\bar\rho$};
\node at (box71.north) [yshift=-1.15cm] {$\rho$};
\node at (box71.north) [yshift=1cm,xshift=.6cm] {MaxMDFT};
\draw[-latex,line width=1pt] ([yshift=.25cm,xshift=.1cm]box51.east) -- ([yshift=.35cm,xshift=-.4cm]box71.west) node[below left] {$\hAA^\Lambda \to \AA=\langle\hAA^\Lambda\rangle$};
\node[box,minimum width=1cm] at (box71.east) [xshift=1.2cm] (box71) {};
\node[boxalt,minimum width=1cm] (box72) at (box71.north) [yshift=-0.1cm,xshift=-.3cm] {};
\node at (box72.north) [yshift=-0.37cm] {$\jj$};
\node at (box72.north) [yshift=-1.1cm,xshift=.1cm] {$\bAA$};
\node at (box71.north) [yshift=-1.15cm] {$\AA$};
\node at ([yshift=.35cm,xshift=-.49cm]box71.west) {$+$};

\end{tikzpicture}
}
\caption{The two discussed boundary cases that arise from entirely removing the field operator $\hAA^\Lambda$ from the description and from a mean-field approximation. While \ac{MaxDFT} is a fully regularized form of \ac{DFT}, \ac{MaxMDFT} also allows to include magnetic fields.}
\label{fig:boundary}
\end{figure}
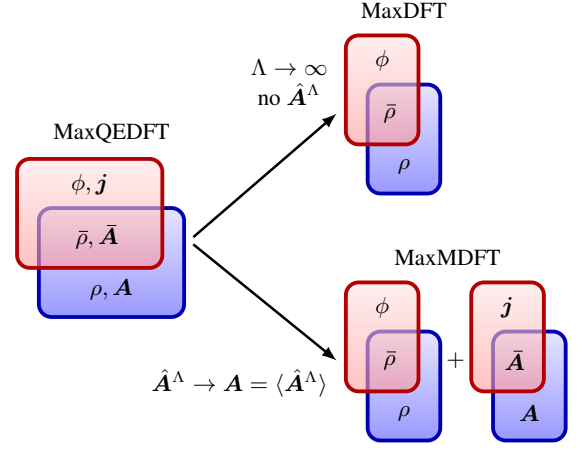

We have kept the cutoff (and the related bare mass $m^\Lambda$) in this mean-field setting, since one encounters similar renormalization issues even in the case of a classical charge coupled to its electromagnetic field~\cite{spohn-book}. While we could perform the same renormalization procedure as before, we here, for simplicity, stay within the $\Lambda$-regularized setting. 
This is almost the setting of \ac{MDFT}~\cite{Tellgren2018-MDFT} (and \emph{not} of \ac{CDFT} as one might have suspected) that also includes the self-energy of the magnetic field and has its own very interesting regularity properties~\cite[Sec.~VI.B]{MY-Perspective}. Yet, one difference to the boundary case of \ac{MaxQEDFT} discussed here remains, and that is that \ac{MDFT} misses the regularization in the $\rho$ component, i.e., it does not switch to $\bar\rho$ and the universal functional misses the energy contribution from the scalar potential $\phi$, and consequently the universal functional is not differentiable in this component.
Despite this difference, the regularized theory that allows to handle magnetic fields without including a quantized photon field developed here can be consistently called \ac{MaxMDFT}.
Since all the results from Sec.~\ref{sec:df-formulation} are preserved in this setting, it offers itself as an alternative to \ac{CDFT}, where no Hohenberg--Kohn theorem can be established~\cite{Penz2023-PartII} and a consistent choice of functional spaces proves difficult~\cite{Laestadius2019}.
Fig.~\ref{fig:boundary} illustrates the two boundary cases discussed in this section.

\section{Discussion and outlook}
\label{sec:discussion}

Let us collect the most relevant results that were obtained before discussing the findings in more detail:

\begin{enumerate}[(i)]

\item We have recast equilibrium non-relativistic \ac{QED} and standard \ac{QM} into a functional-theoretic formulation based on Hilbert spaces and in terms of mixed (combined external and internal) charge densities and currents that includes the field energy of the electromagnetic environment into the universal functional.

\item We have made \ac{QEDFT} and its sibling theories, standard ground-state \ac{DFT} and \ac{MDFT}, Fr\'echet differentiable and can thus guarantee the unique existence of a Kohn--Sham systems. We call those theories `Maxwell-regularized' and use the acronyms \ac{MaxQEDFT}, \ac{MaxDFT}, and \ac{MaxMDFT}.

\item We found that the main variables of the new theories correspond to the bare sources of the electromagnetic environment that allows for back-reactions from the quantum system and thus couples non-linearly. This builds a bridge to the macroscopic Maxwell equations that suggest a natural identification of the involved quantities.

\item We have removed the ultra-violet cutoff from Pauli--Fierz quantum field theory on the level of a functional theory and showed the existence of a (potentially multi-phase) non-linear renormalized Kohn--Sham theory in terms of the mixed quantities.

\item We find the independent sibling theories with magnetic fields but without field quantization (\ac{MaxMDFT}) and the density-only functional theory (\ac{MaxDFT}) as exact boundary cases of \ac{MaxQEDFT} that share the same mathematical framework.

\end{enumerate}
All of this was possible with the following techniques and principles:

\begin{enumerate}[(a)]

\item We have included a confining potential that breaks the symmetry, sets an overall scale, localizes the system, and guarantees existence of a ground state.
This allows to choose a bound-state reference system instead of the usual scattering-state reference.

\item We treat the external sources not simply as parameters in the Hamiltonian, but acknowledged their physical existence and included their intrinsic energy into the universal functional.

\item We have chosen the sources from spaces that already embed gauge freedom, and that together with their dual spaces for densities and vector potentials encode the Maxwell equations geometrically by employing energy norms. We also carefully showed that these spaces fit to the Hamiltonian, in the sense that they guarantee its self-adjointness on a stable domain, and that they include the usual setting of \ac{DFT}.

\item We have considered \emph{mixed} charge densities and vector potentials that result from a combination of the internal quantities and the external sources. As such, a mixed density can take arbitrary (positive and negative) values and becomes part of the same space as the external densities that enter the Maxwell equations. This works similarly for the internal vector potential and external currents.

\item In the corresponding Kohn--Sham construction the non-interacting non-coupled system is chosen such that it produces the mixed quantities instead of the usual internal ones. This circumvents all $v$-representability problems and leads to a well-defined Maxwell-regularized \ac{Hxc} potential that also takes an important role in removing the ultra-violet cutoff. Through the interpretation of the mixed quantities as the bare sources, this leads to the insight that we consider two different systems in the same electromagnetic environment in this construction.

\end{enumerate}
The resulting universal functional suggests that what we have actually done physically is to recast the entire linear quantum theory as a non-linear theory for the sources, which themselves react to the quantum system coupled to them. As such $(-\bar \rho, \bAA)^*$ can be interpreted as bare classical source terms that get screened by the quantum system. Those bare sources are \emph{universal} in the sense that they do not depend on the specific (quantum) system coupled to them. The interaction between system and environment self-consistently leads to $(\phi,\jj)$ that one usually considers as parameters of the Hamiltonian and that then allow to determine the charge density and magnetic field of the quantum system from the ground state. Since we have promoted the \emph{universal bare sources} to the basic quantities of our description, it becomes possible to easily compare different systems and perform limiting procedures. All the details of the quantum system can in principle be encapsulated in gauge-independent macroscopic Maxwell equations for the coupled quantum-classical system and an observer-theoretic perspective is natural. We will derive and discuss Maxwell-regularized \ac{QED} from this physically motivated perspective in a separate publication. In this future work we also plan to switch to a periodic setting to highlight that the derived approach is completely general and does not rely on the chosen confining potential or corresponding reference solution. Still, for completeness, let us clarify already here how Maxwell-regularized \ac{QED} naturally arises also in the periodic case. This is the standard setting for solid-state systems and without the need for a confining potential we set $\phiconf =0$. The reference density is then $\rhoref = -e \, N/V$ and we thus consider densities relative to this homogeneous solution. By construction, the theory then does not contain the non-physical $\kk =0$ contributions and fits exactly to the usual construction of the local-density approximation~\cite{dreizler-gross-book}.

In such a periodic setting, we can straightforwardly include also the nuclei as quantized particles, since we have a purely discrete spectrum in this case by construction. The classical densities and fields are then left with the role of classical trapping and probing fields, i.e., the details of the experimental setup according to a Heisenberg cut. Thus the Born--Oppenheimer-type setting discussed in this manuscript is mainly chosen for convenience. But it also allows to highlight the conceptual shift that has happened in this work: We do not ask for control/knowledge of an (artificial/arbitrary) subsystem, which depends on the Born--Oppenheimer-type approximation, the ultra-violet regularization, and the gauge choice, but instead we focus on the \emph{mixed} density and vector potential. The price we pay is that we have a highly implicit and non-linear theory. In other words, wave functions and density matrices are reduced to convenient but purely theoretical concepts of theory building for a non-linear theory in terms of only physical (directly measurable) objects. 

This conceptual shift allows to remove the ultra-violet cutoff from the Pauli--Fierz theory. While on a first glance it seems inconvenient that we can have multiple limits in Eq.~\eqref{eq:phiLambdaLimit}, this is a quite natural result. Following the usual Gelfand--Naimark--Segal construction~\cite{thirring2002quantum} for different types of limits, e.g., thermodynamical or also classical limits, we naturally end up with potentially different phases \emph{outside} of the original separable Hilbert-space structure~\cite{Ruetsche_2003,van2023emergent,van2024gibbs}. In practice, standard ground-state \ac{DFT} employs this connection to an emerging novel structure very successfully since many years, for instance, when the \ac{LDA} derived from an infinite system is applied as an approximation to the \ac{Hxc} energy of a finite system~\cite{dreizler-gross-book}.

On the side of \ac{DFT} and \ac{QEDFT} we see that we do not aim at making the internal densities and vector potentials the same between the coupled system and the auxiliary Kohn--Sham system any more, but instead only match the mixed densities and vector potentials. We could now develop custom-built approximation that take into account the explicit connection to the Maxwell theory, which allows to use new physical ideas even for ground-state \ac{DFT}. 
It is to be expected that the explicit variational form of the Maxwell-regularized universal functional from Eq.~\eqref{eq:Fbar-variational} will provide a valuable stepping stone for this task.
Moreover, since the functionals in the regularized theory are now Fr\'echet differentiable and hence vary smoothly, there is hope that approximations might become more controlled in this new framework. Finally, for finite-dimensional Hilbert spaces, one can even show convergence of the regularized Kohn--Sham iteration~\cite{KS_PRL_2019,KS_PRL_2019_Erratum} and it is anticipated that a similar proof extends to the present infinite-dimensional Hilbert space setting.

As a further outlook, let us note that after having renormalized non-relativistic \ac{QED} of equilibrium systems in the form of a functional theory, it seems entirely possible to do the same for second-quantized Dirac theory. We could take \ac{QED} Hamiltonians with cutoffs of the form discussed by Ref.~\cite{10.1063/1.3133885} and apply an analogous strategy. Even more interesting might be the connection to other scales and theories.
Instead of electromagnetic fields we could consider also non-abelian gauge theories, where a matter current couples minimally to the generators of a Lie group and the exact same structure arises. The Maxwell-regularization is then replaced by energy terms from a classical Yang--Mills field theory.
Since the wave function has effectively vanished from our theory (nicely encapsulated in the universal functional) and only physical densities and fields appear, it seems possible to connect the remaining fundamental quantities straightforwardly to macroscopic theories. Indeed, we have reformulated ground-state quantum physics in a rigorous manner as a highly non-linear theory connecting to the Maxwell equations and this brings it structurally already much closer to general relativity. In rough terms, instead of making gravity quantum, we made quantum more gravity. Another important question is the time-dependent generalization of this approach. \ac{QEDFT} and \ac{DFT} allow an (at least formally) straightforward formulation in terms of equations of motions~\cite{MarquesMaitraNogueiraGrossRubio2012,ruggenthaler2015existence,ruggenthaler2014quantum,ruggenthaler2023understanding}. A similar formulation for \ac{MaxQEDFT} and \ac{MaxDFT} seems possible but still needs to be developed in detail. That a time-dependent formulation is desirable becomes most obvious in the \ac{QED} setting. If we work with the fully quantized electromagnetic field, the theory is purely local. This is in contrast to equilibrium \ac{QM}, which is the infinite-time limit of \ac{QED} and thus non-local~\cite{ruggenthaler2023understanding}. Despite still building upon a self-adjoint Hamiltonian, \ac{QED} becomes \emph{effectively dissipative}, since the system will relax towards its ground state upon emitting photons into the far field that offers a continuum of eigenvalues, a process that is termed spontaneous emission~\cite[Sec.~17.2]{spohn-book}. So from this point of view, a time-dependent \ac{QEDFT} calculation would only feature \emph{local} interactions (for which a Kohn--Sham description seems natural) and upon propagating long enough the system will always locally converge to the ground state. In other words, in such a time-dependent formulation that includes the (quantized) Maxwell field explicitly, it seems possible to exchange spatial non-locality and entanglement with explicit time-evolution.

\newcommand{\nocontentsline}[3]{}
\let\origaddcontentsline\addcontentsline
\newcommand{\stoptoc}{\let\addcontentsline\nocontentsline}
\newcommand{\resumetoc}{\let\addcontentsline\origaddcontentsline}
\stoptoc

\section*{Acronyms}
\begin{acronym}
\setlength{\itemsep}{-1.7em} 

\acro{DFA}{density-functional approximations}
\acro{DFT}{density-functional theory}
\acro{CDFT}{current-density-functional theory}
\acro{GEA}{gradient expansion approximation}
\acro{GGA}{generalized gradient approximation}
\acro{Hxc}{Hartree-exchange-correlation}
\acro{LDA}{local-density approximation}
\acro{MaxDFT}{Maxwell-regularized density-functional theory}
\acro{MaxMDFT}{Maxwell-regularized Maxwell--Schr\"{o}dinger density-functional theory}
\acro{MaxQEDFT}{Maxwell-regularized quantum-electrodynamical functional theory}
\acro{MDFT}{Maxwell--Schr\"{o}dinger density-functional theory}
\acro{OEP}{optimized effective potential}
\acro{QED}{quantum electrodynamics}
\acro{QEDFT}{quantum-electrodynamical functional theory}
\acro{QM}{quantum mechanics}
\acro{RPA}{random phase approximation}
\end{acronym}

\section*{AI Disclaimer}
This work was conducted entirely by the authors. No artificial intelligence (AI) tools were used in the derivation, analysis, or validation of any scientific results presented in this paper. However, AI tools were used to help identify relevant references, to do background research, and in a few instances for language formatting. All content, interpretations, and conclusions are the sole responsibility of the authors.

\section*{Data availability}
No new data were created or analyzed in support of this research.

\acknowledgments

The idea for Maxwell regularization in DFT can be traced back to talks given by Trygve Helgaker in 2012 and the work on Maxwell--Schr\"odinger DFT by Erik I.\ Tellgren. Moreover, Simen Kvaal and Andre Laestadius took an important role in its early inception and long-term development. We thank them for continued exchange on the topic and hope that this theoretical offspring lives up to their original vision. Further, we are grateful for inspiring discussions with Nicolas Tancogne-Dejean.
Funding for MP is acknowledged from the German Research Foundation under the grant SCHI 1476/1-1 and from the European Research Council under the grant ERC-2021-STG No.~101041487 REGAL.
CJ, MR, and AR acknowledge support by the European Research Council under the grant ERC-2024-SyG No.~101167294  UnMySt. The views and opinions expressed in this work are, however, those of the authors only and do not necessarily reflect those of the European Union or the European Research Council. Neither the European Union nor the European Research Council can be held responsible for them. CJ, MR, and AR further acknowledge support from the Cluster of Excellence Advanced Imaging of Matter (AIM), Grupos Consolidados (IT1249-19), and SFB925.

\bibliography{refs}
\end{document}